\documentclass[journal]{IEEEtran}
\makeatletter

\newcommand{\Rmnum}[1]{\expandafter\@slowromancap\romannumeral #1@}
\makeatother
\usepackage{amsmath,amsfonts}
\usepackage{algorithmic}
\usepackage{algorithm}
\usepackage{array}
\usepackage[caption=false,font=normalsize,labelfont=sf,textfont=sf]{subfig}
\usepackage{textcomp}
\usepackage{stfloats}
\usepackage{url}
\usepackage{verbatim}
\usepackage{graphicx}
\usepackage{multirow}
\usepackage{cite}
\usepackage{booktabs}
\usepackage{siunitx}
\usepackage{hyperref}
\usepackage[export]{adjustbox}
\usepackage{lipsum}
\usepackage{comment}
\usepackage[switch]{lineno}
\begin{document}

\title{A Fast and Accurate 3-D Reconstruction Algorithm for Near-Range Microwave Imaging with Handheld Synthetic Aperture Radar}

\author{Lei Wang, Xianxun Yao, Tiancheng Song, and Guolin Sun
\thanks{The authors are with the School of Electronic and Information Engineering, Beihang University, Beijing 100191, China (e-mail: \href{mailto:xianxun.yao@buaa.edu.cn}{\mbox{xianxun.yao@buaa.edu.cn}}).} 
\thanks{Manuscript received November 1, 2024.

This work has been submitted to the IEEE for possible publication. Copyright may be transferred without notice, after which this version may no longer be accessible.}}

\markboth{IEEE...,~Vol.~XX, No.~XX, XXXX~XXXX}
{ }

\IEEEpubid{0000--0000/00\$00.00~\copyright~2021 IEEE}
\maketitle

\begin{abstract}
The design of image reconstruction algorithms for near-range handheld synthetic aperture radar (SAR) systems has gained increasing popularity due to the promising performance of portable millimeter-wave (MMW) imaging devices in various application fields. Time domain imaging algorithms including the backprojection algorithm (BPA) and the Kirchhoff migration algorithm (KMA) are widely adopted due to their direct applicability to arbitrary scan trajectories. However, they suffer from time complexity issues that hinder their practical application. Wavenumber domain algorithms greatly improve the computational efficiency but most of them are restricted to specific array topologies. Based on the factorization techniques as adopted in far-field synthetic aperture radar imaging, the time domain fast factorized backprojection algorithm for handheld synthetic aperture radar (HHFFBPA) is proposed. The local spectral properties of the radar images for handheld systems are analyzed and analytical spectrum compression techniques are derived to realize efficient sampling of the subimages. Validated through numerical simulations and experiments, HHFFBPA achieves fast and accurate 3-D imaging for handheld synthetic aperture radar systems with arbitrary trajectories.
\end{abstract}

\begin{IEEEkeywords}
millimeter-wave (MMW) imaging, 
3-D near-range imaging, 
synthetic aperture radar (SAR), 
handheld SAR imaging,
fast factorized backprojection (FFBP).
\end{IEEEkeywords}

\section{Introduction}

\IEEEPARstart{N}{ear-range} millimeter-wave 3-D imaging techniques \cite{lopez20003,942570} have been extensively researched and applied in various fields, including nondestructive testing and evaluation (NDT\&E) \cite{8007280,kharkovsky2007microwave}, medical diagnosis \cite{bolomey1990microwave,klemm2010microwave}, through-wall imaging \cite{4454449,chang2009adaptive}, and security screening \cite{wang2024elliptical,song2024fast,wang2024space}, because of its penetrative imaging abilities, non-ionizing properties and robustness against lighting conditions.

\IEEEpubidadjcol

The imaging resolution of array-based millimeter-wave 3-D imaging systems primarily depends on the size of the array aperture and the working bandwidth, and millimeter resolutions can be achieved by employing a large aperture and wide working bandwidth. Conventional 3-D millimeter-wave imaging systems are primarily implemented through two methods. The first method involves electronically scanned 2-D single-input-single-output (SISO) arrays or multiple-input-multiple-output (MIMO) arrays, where a high number of physical array elements are arranged on a 2-D aperture, like the fully electronic active real-time imager developed by R\&S \cite{6080744,ahmed2012advanced}. By employing electronic switching of the array elements, the scattered field of the targets at different positions can be measured, enabling high-speed data acquisition. However, this approach incurs significant costs due to the requirement of numerous physical array elements. Alternatively, a more cost-effective method is using mechanically scanned synthetic aperture arrays \cite{9136646,5530374,6015544}, as demonstrated by the Pacific Northwest National Laboratory (PNNL) \cite{942570}. In this case, the physical array with fewer array elements is mechanically scanned to form a large synthetic aperture. To achieve high resolutions, large-scale mechanical scanning hardware systems are required to obtain large synthetic apertures.

\IEEEpubidadjcol

However, in many applications including nondestructive testing and security screening, the primary requirements for 3-D millimeter-wave imaging systems revolve around the need for compactness, portability, and affordability. Consequently, handheld SAR imaging systems \cite{ghasr2011portable,ghasr2016wideband,alvarez2019freehand,laviada2017multiview,he2016development,alvarez2022freehand,li2024ifnet} have been proposed and extensively researched as a competitive solution where the physical array is scanned manually. Contrary to conventional SAR systems, the positions of the array elements in handheld SAR systems exhibit significant uncertainty and non-uniformity, due to the inherent instability of manual scanning. This trajectory uncertainty greatly increased the complexity of image reconstruction algorithms. The main challenges include precise motion tracking of the radar system and the development of fast and accurate 3-D imaging algorithms tailored for handheld SAR systems with arbitrary trajectories.

Numerous related studies have been published. The basic approach involves employing a high-precision motion tracking system \cite{alvarez2020freehand,ghasr2011portable,alvarez2021freehand} to accurately capture the motion trajectory and orientation of the physical array, albeit at the expense of increased system complexity and cost. To mitigate expenses and further improve the accuracy of the positioning system, various auxiliary positioning methods have been introduced and employed, including introducing reference targets and performing online calibrations \cite{li2024high,alvarez2021towards} or through iterative optimization techniques \cite{schellberg2023mmsight}. Furthermore, several data-driven autofocus methods have also been proposed. By adjusting the phase error of the received signals caused by position deviations of array elements, imaging quality indicators such as entropy \cite{zeng2013sar,wang2006sar} and sharpness \cite{morrison2007sar,fienup2000synthetic} can be numerically optimized and the imaging results can exhibit high consistency with the ground truth \cite{laviada2022artifact}.

Among the aforementioned methods, extensive research has been conducted on precise motion tracking of the radar system. However, the problem of designing fast and accurate 3-D imaging algorithms tailored for handheld SAR systems is less discussed, which constitutes the main focus of this paper.

Time domain imaging algorithms, such as the back projection algorithm (BPA) \cite{kazemi2018spectral} and the Kirchhoff migration algorithm (KMA) \cite{5422639}, can effectively compensate for the trajectory deviations at each scan position to achieve high-precision image reconstruction. These algorithms are directly applicable to non-uniform synthetic arrays and are widely employed in handheld imaging applications. However, their practical implementation is limited by the significant computational complexity, hindering their applications in real-time imaging systems.

In order to alleviate the computational burden of the time domain algorithms, several efficient wavenumber domain imaging algorithms \cite{8094359,9126860,9082116} have been proposed, which utilize spherical wave decomposition, phase shift migration, and other techniques. However, most of these algorithms are exclusively designed for uniform arrays. For non-uniform arrays, the range migration algorithm based on non-uniform fast Fourier transform (NUFFT-based RMA) \cite{8974215,chen2022efficient} is introduced. Nevertheless, wavenumber domain algorithms are often restricted to planar arrays, cylindrical arrays, and other arrays with well-defined regular trajectories. Alternatively, the equivalent phase center (EPC) approximation-based range migration algorithm (EPC-RMA) \cite{9690108,7832573} is proposed where the measured data from non-uniform arrays with arbitrary trajectory can be approximately calibrated onto a 2-D planar uniform array and conventional range migration algorithms can be used. The phase errors caused by such approximation cause imaging quality degradation which is prohibitive for various applications requiring precise imaging. However, due to its high computational efficiency, it remains widely adopted in handheld SAR imaging.

In this paper, a fast yet accurate 3-D imaging algorithm for near-range handheld SAR systems is proposed. Considering that the analytical dispersion relation in the wavenumber domain for non-uniform arrays with arbitrary trajectory is hard to derive, the proposed algorithm is designed in the time domain. The factorization technique, which was first introduced in the fast backprojection algorithm (FBPA) \cite{767270,shao2013fast} and the fast factorized backprojection algorithm (FFBPA) \cite{1238734,zhou2017quasi} for far-field SAR imaging, is adopted to lower the computational complexity while preserving the applicability to synthetic arrays with arbitrary trajectory. The complete synthetic array is partitioned into subarrays with multiple levels. Then, downsampled subimages of each subarray are reconstructed with time domain imaging algorithms for the first level and merged from lower-level subimages for the subsequent levels.

The achievable computational efficiency is determined by the sampling point reduction ratio of the subimages. Unlike the far-field SAR settings where the spectrum properties of the radar images are simple and heuristic spectrum compression techniques can be directly designed, the spectrum properties of near-range radar images are strongly space-variant. \cite{song2024fast} introduced the fast factorized Kirchhoff migration algorithm (FFKMA) for near-range MIMO systems. The local spectral properties of the near-range radar images are analyzed and an efficient spectrum compression method including spatial downconversion (SDC) and local linear transforms (LLT) is established. However, this algorithm requires complex numerical operations including global optimizations and discrete Helmholtz-Hodge decomposition operations to calculate the interpolation indices and other information as a precalculation step for a given array. This is sub-optimal for handheld systems where the array configuration varies for every manual scan. Therefore, the proposed algorithm eliminates the complex numerical precalculation steps by deriving analytical forms of SDC and LLT specifically for handheld systems. As a result, the proposed algorithm, named handheld fast factorized backprojection algorithm (HHFFBPA), achieves accurate and efficient 3-D imaging for handheld SAR imaging with arbitrary trajectory.

This article is organized as follows. In Section \ref{sec_sgnmod}, the signal model of the handheld SAR imaging system, the formulation of BPA, and the analytical local spectral properties of the imaging results are given. In Section \ref{sec_ffbp}, the design and formulation of the proposed HHFFBPA are described. In Section \ref{sec_impcmp}, the implementation and complexity analysis of the proposed algorithm are discussed in detail. In Section \ref{sec_simexp}, the proposed algorithm is validated with numerical simulations and experiments and is compared with other imaging algorithms. Finally, the conclusion is summarized in Section \ref{sec_conlu}.

\section{Signal Model}\label{sec_sgnmod}

\subsection{Handheld SAR Imaging System}

\begin{figure}[htbp]\centering
\includegraphics[width=3.5in]{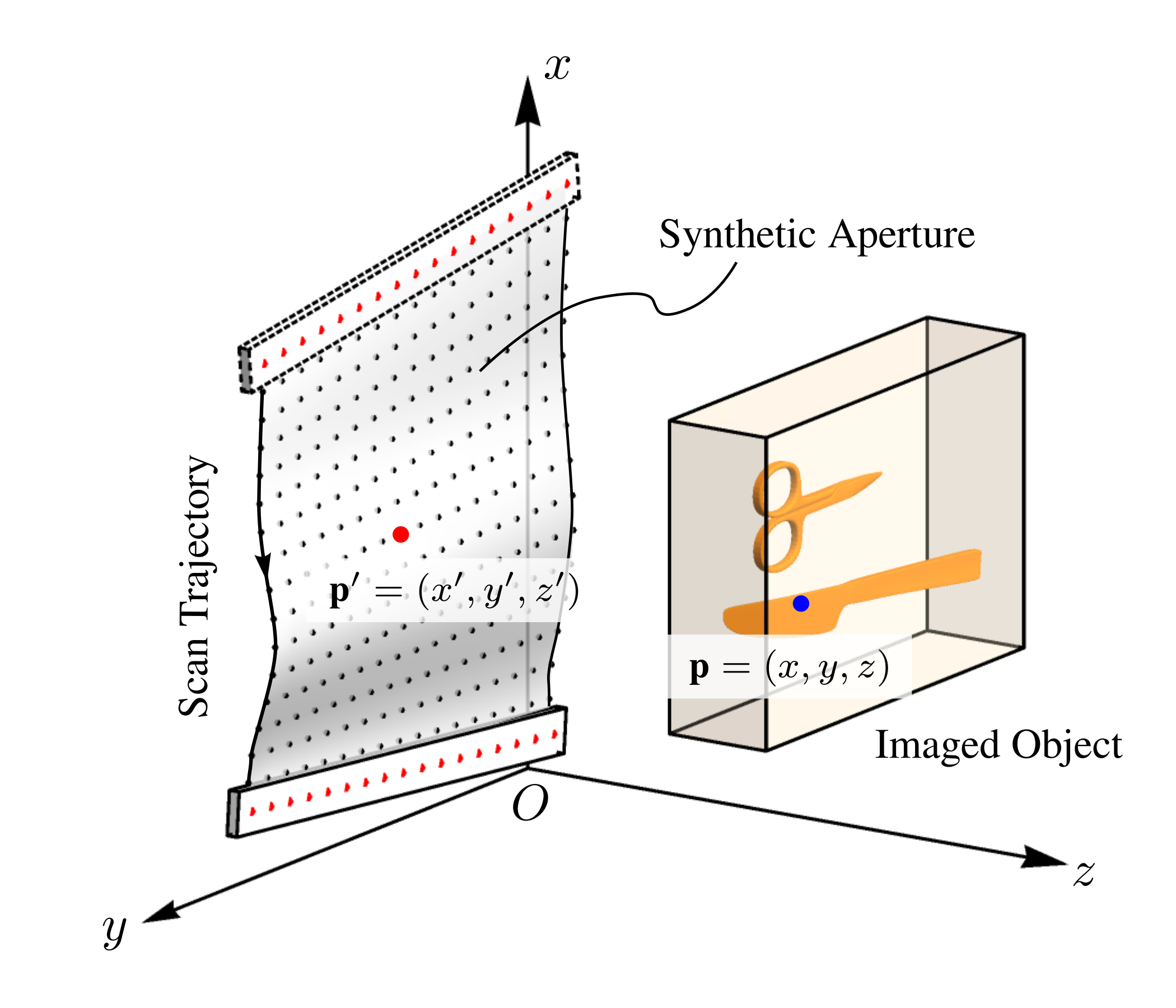}
\caption{Geometry of near-range handheld SAR imaging systems.}
\label{fig_scenario}
\end{figure}

The geometry of near-range handheld SAR imaging systems is depicted in Fig. \ref{fig_scenario}. The radar array, like the linear single-input-single-output (SISO) array shown in the figure as an example, is manually scanned to form a synthetic aperture and illuminates the imaged objects located in front of the aperture. In contrast to traditional SAR systems with rigid mechanical scanners, handheld systems exhibit significant uncertainty on the scan trajectory, resulting in fluctuations in the synthesized array element positions.

The coordinates of the array elements within the synthetic aperture are denoted as $\textbf{p}'=(x',y',z')$, while the coordinates of the scattering points in the imaging region are denoted as $\textbf{p}=(x,y,z)$. By employing the first-order Born approximation and neglecting the propagation losses, the measured signal can be represented as
\begin{equation}\label{}
\begin{aligned}
s(\textbf{p}',k) = \iiint\limits_{D}f(\textbf{p})e^{-j2k||\textbf{p}-\textbf{p}'||}d\textbf{p},
\end{aligned}
\end{equation}
where $k=2{\pi}f/c$ is the wavenumber of the MMW signal. $c$ is the speed of light and $f$ is the frequency of the MMW signal. $||\bullet||$ denotes the magnitude of the vector and $||\textbf{p}-\textbf{p}'||$ denotes the distance between the array element and the scattering point. $D$ represents the imaging region encompassing the imaged targets. $f(\textbf{p})$ represents the reflectivity function of the target, which is the objective of the image reconstruction process.

Among the existing image reconstruction algorithms, BPA is widely acknowledged as the gold standard due to its exceptional imaging quality. The measured signal is first transformed into the time domain using the fast inverse Fourier transform and then reconstructed using BPA, which is formulated as
\begin{equation}\label{eq2_rangecompression}
\begin{aligned}
s(\textbf{p}',t)
&=\int_{k_{\min}}^{k_{\max}}s(\textbf{p}',k)e^{jkct}dk,
\end{aligned}
\end{equation}
\begin{equation}\label{eq2_bpatimedomain}
\begin{aligned}
f(\textbf{p}) = \iint\limits_As(\textbf{p}',\tau)|_{\tau=\frac{2||\textbf{p}-\textbf{p}'||}{c}}d\textbf{p}',
\end{aligned}
\end{equation}
where $k_{\min}$ and $k_{\max}$ represent the lower and upper bounds of the wavenumber of the MMW signal, respectively. $A$ represents the synthetic aperture, and $\tau$ denotes the round-trip propagation delay between the array element and the scattering point in the imaging region.

By substituting \eqref{eq2_rangecompression} into \eqref{eq2_bpatimedomain}, the overall formulation of BPA is
\begin{equation}\label{eq2_bpafreqdomain}
\begin{aligned}
f(\textbf{p}) = \iint\limits_A\int_{k_{\min}}^{k_{\max}}s(\textbf{p}',k)e^{j2k||\textbf{p}-\textbf{p}'||}d\textbf{p}'dk.
\end{aligned}
\end{equation}

In (\ref{eq2_bpafreqdomain}), BPA reconstructs the imaging result on a pixel-by-pixel basis by performing the spatial-wavenumber integration at each position within $D$. Therefore, the computational load of BPA is proportional to the number of sampling points of the imaging result, which is determined by the spectral bandwidth properties of the imaging result. For near-range SAR images, the spectral properties including the carrier frequencies and bandwidths are strongly space-variant. As a result, local spectrum analysis techniques are performed to better model the spectral properties of the handheld SAR images.

\subsection{Local Spectral Properties of the Handheld SAR Images} \label{subseclsp}

The wavenumber domain representation of the handheld SAR images can be straightforwardly calculated through the 3-D Fourier transform, which is
\begin{equation}\label{}
\begin{aligned}
F(\textbf{k}) 
&=\iiint\limits_{D}f(\textbf{p})e^{-j\textbf{k}\cdot\textbf{p}}d\textbf{p} \\
&=\iiint\limits_D\iint\limits_A\int_{k_{\min}}^{k_{\max}}s(\textbf{p}',k)e^{j2k||\textbf{p}-\textbf{p}'||-j\textbf{k}\cdot\textbf{p}}d\textbf{p}d\textbf{p}'dk,
\end{aligned}
\end{equation}
where $\textbf{k}=(k_x,k_y,k_z)$ represents the wavenumber domain variable of the image spectrum. $k_x$, $k_y$, and $k_z$ represent the wavenumber domain counterparts of $x$, $y$, and $z$ respectively.

The principle of stationary phase (POSP) is then utilized to solve the above integral over $x,y,z$, which is formulated as
\begin{equation}\label{eq2_imgspectrumposp}
\begin{aligned}
F(\textbf{k})
&=\iint\limits_{A}\int_{k_{\min}}^{k_{\max}}{\alpha}s(\textbf{p}',k)e^{j2k||\textbf{p}_0-\textbf{p}'||-j\textbf{k}\cdot\textbf{p}_0}d\textbf{p}'dk,
\end{aligned}
\end{equation}
where $\alpha$ and $\textbf{p}_0=(x_0,y_0,z_0)$ denote the slow-varying amplitude term and the stationary point in POSP, respectively. 

The stationary point satisfies
\begin{equation}\label{}
\begin{aligned}
\nabla(2k||\textbf{p}-\textbf{p}'||-\textbf{k}\cdot\textbf{p})|_{\textbf{p}=\textbf{p}_0}=\textbf{0},
\end{aligned}
\end{equation}
which is further simplified as
\begin{equation}\label{eq_k_posp}
\begin{aligned}
2k\nabla(||\textbf{p}-\textbf{p}'||)|_{\textbf{p}=\textbf{p}_0}=\textbf{k}.
\end{aligned}
\end{equation}

Therefore, the expression of any $\textbf{k}=\textbf{k}_0=(k_x,k_y,k_z)$ that satisfies (\ref{eq_k_posp}) is 
\begin{equation}\label{eq_kxyz_posp}
\begin{aligned}
k_x&=2k\frac{x_0-x'}{\sqrt{(x_0-x')^2+(y_0-y')^2+(z_0-z')^2}}\\
k_y&=2k\frac{y_0-y'}{\sqrt{(x_0-x')^2+(y_0-y')^2+(z_0-z')^2}}\\
k_z&=2k\frac{z_0-z'}{\sqrt{(x_0-x')^2+(y_0-y')^2+(z_0-z')^2}}.
\end{aligned}
\end{equation}

Alternatively, $\textbf{k}_0$ can be denoted as a function of $\textbf{p}'$, $k$ and $\textbf{p}_0$
\begin{equation}\label{eq_kfun_posp}
\begin{aligned}
\textbf{k}_0(\textbf{p}',\textbf{p}_0,k)=2k\nabla(||\textbf{p}-\textbf{p}'||)|_{\textbf{p}=\textbf{p}_0}.
\end{aligned}
\end{equation}

The relationship between wavenumber domain variables and system parameter variables is established by equations \eqref{eq_kxyz_posp} and \eqref{eq_kfun_posp}. For any wavenumber domain variables $k_x$, $k_y$, $k_z$, the asymptotic integration result of \eqref{eq2_imgspectrumposp} is non-zero only when \eqref{eq_k_posp} and \eqref{eq_kxyz_posp} are satisfied, which corresponds to a set of limited number of wavenumber domain points. This property can be utilized to determine the spectrum support range of the imaging results, which is formulated as
\begin{equation} \label{eq_sp}
\begin{aligned}
SP_{A} = \{\textbf{k}| 
&\exists\ \textbf{p}_0\in D,\textbf{p}'\in A,k\in [k_{\min},k_{\max}] \\
&\mathrm{s.t.}\ 2k\nabla(||\textbf{p}-\textbf{p}'||)|_{\textbf{p}=\textbf{p}_0} = \textbf{k} \}.
\end{aligned}
\end{equation}

For an arbitrary position in the imaging area, $\textbf{p}_0 \in D$, the corresponding spectrum support range can be calculated through \eqref{eq_k_posp} and \eqref{eq_kxyz_posp}:
\begin{equation}\label{}
\begin{aligned}
SP_{A,\textbf{p}_0}=\{\textbf{k}|
&\exists\ \textbf{p}'\in A,k\in [k_{\min},k_{\max}]\\
&\mathrm{s.t.}\ 2k\nabla(||\textbf{p}-\textbf{p}'||)|_{\textbf{p}=\textbf{p}_0}=\textbf{k}\},
\end{aligned}
\end{equation}
which is defined as the local spectrum support at $\textbf{p}_0$.

Accordingly, the spectrum support of the global image can be represented as the union of the local spectrum supports at each position in the imaging region, that is
\begin{equation}\label{eq_spcm_locspcm_union}
\begin{aligned}
SP_{A}=\bigcup\limits_{\textbf{p}_0\in D}SP_{A,\textbf{p}_0}.
\end{aligned}
\end{equation}

Equation \eqref{eq_sp} to \eqref{eq_spcm_locspcm_union} describes the properties of handheld SAR image spectrum support. The Nyquist sampling rates along each dimension can now be derived accordingly, which are
\begin{equation}\label{eq_dxyz_kxyzminmax}
\begin{aligned}
&\Delta{x}=\frac{2\pi}{k_{x\max}-k_{x\min}}\\
&\Delta{y}=\frac{2\pi}{k_{y\max}-k_{y\min}}\\
&\Delta{z}=\frac{2\pi}{k_{z\max}-k_{z\min}},
\end{aligned}
\end{equation}
where $k_{x\min}$, $k_{x\max}$, $k_{y\min}$, $k_{y\max}$, $k_{z\min}$, and $k_{z\max}$ represent the lower and upper bounds of the $k_x$, $k_y$, and $k_z$ determined by \eqref{eq_sp}, respectively.

Ignoring the minor fluctuations on the $z$-axis of the array elements for simplicity of the formulations, the above formula can be further expanded as
\begin{equation}\label{eq_dxyz_xyzminmax}
\begin{aligned}
\Delta{x}\approx&\frac{\pi}{k_{\max}(\frac{x_{\max}-x'_{\min}}{\sqrt{(x_{\max}-x'_{\min})^2+(z_{\min})^2}}-\frac{x_{\min}-x'_{\max}}{\sqrt{(x_{\min}-x'_{\max})^2+(z_{\min})^2}})}\\
\Delta{y}\approx&\frac{\pi}{k_{\max}(\frac{y_{\max}-y_{\min}}{\sqrt{(y_{\max}-y'_{\min})^2+(z_{\min})^2}}-\frac{y_{\min}-y'_{\max}}{\sqrt{(y_{\min}-y'_{\max})^2+(z_{\min})^2}})}\\
\Delta{z}\approx&\frac{\pi}{k_{\max}-k_{\min}\frac{z_{\min}}{\sqrt{(x_{d,\max})^2+(y_{d,\max})^2+(z_{\min})^2}}},
\end{aligned}
\end{equation}
where $x_{\min}$, $x_{\max}$, $y_{\min}$, $y_{\max}$, and $z_{\min}$ denote the lower and upper bounds of $x$, $y$, and the lower bound of $z$ of the imaging region. $x'_{\min}$, $x'_{\max}$, $y'_{\min}$, $y'_{\max}$ denote the lower and upper bounds of $x'$ and $y'$ within the synthetic aperture. $x_{d,\max}$ and $y_{d,\max}$ represent the maximum of $|x-x'|$ and $|y-y'|$, which is formulated as follows:
\begin{equation}\label{}
\begin{aligned}
x_{d,\max} =& \begin{cases}
    |x_{\min}-x'_{\max}|,&{\text{if}}\ |x_{\min}-x'_{\max}|>|x_{\max}-x'_{\min}|\\ 
    |x_{\max}-x'_{\min}|,&{\text{otherwise}} 
\end{cases}\\
y_{d,\max} =& \begin{cases}
    |y_{\min}-y'_{\max}|,&{\text{if}}\ |y_{\min}-y'_{\max}|>|y_{\max}-y'_{\min}|\\ 
    |y_{\max}-y'_{\min}|,&{\text{otherwise}}.
\end{cases}
\end{aligned}
\end{equation}

Finally, the number of required sampling points for the imaging results is
\begin{equation}\label{}
\begin{aligned}
N_{s,A}=\iiint\limits_{D}\frac{1}{\Delta{x}\Delta{y}\Delta{z}}dxdydz=\frac{V(D)}{\Delta{x}\Delta{y}\Delta{z}},
\end{aligned}
\end{equation}
where $V(D)$ is the volume of the imaging region $D$.

From the discussions above, it is revealed that the handheld SAR images can be modeled as space-variant band-limited 3-D signals, and the spectrum supports of the images at arbitrary positions are analytically derived. This enables the design of efficient spectrum compression methods combined with the factorization technique to reduce the sampling requirements and therefore the computational load of the time domain reconstruction algorithms.

\section{Fast Factorized Backprojection Algorithm for Near-Range Handheld SAR Imaging}\label{sec_ffbp}

\begin{figure}[htbp]\centering
\includegraphics[width=3.5in]{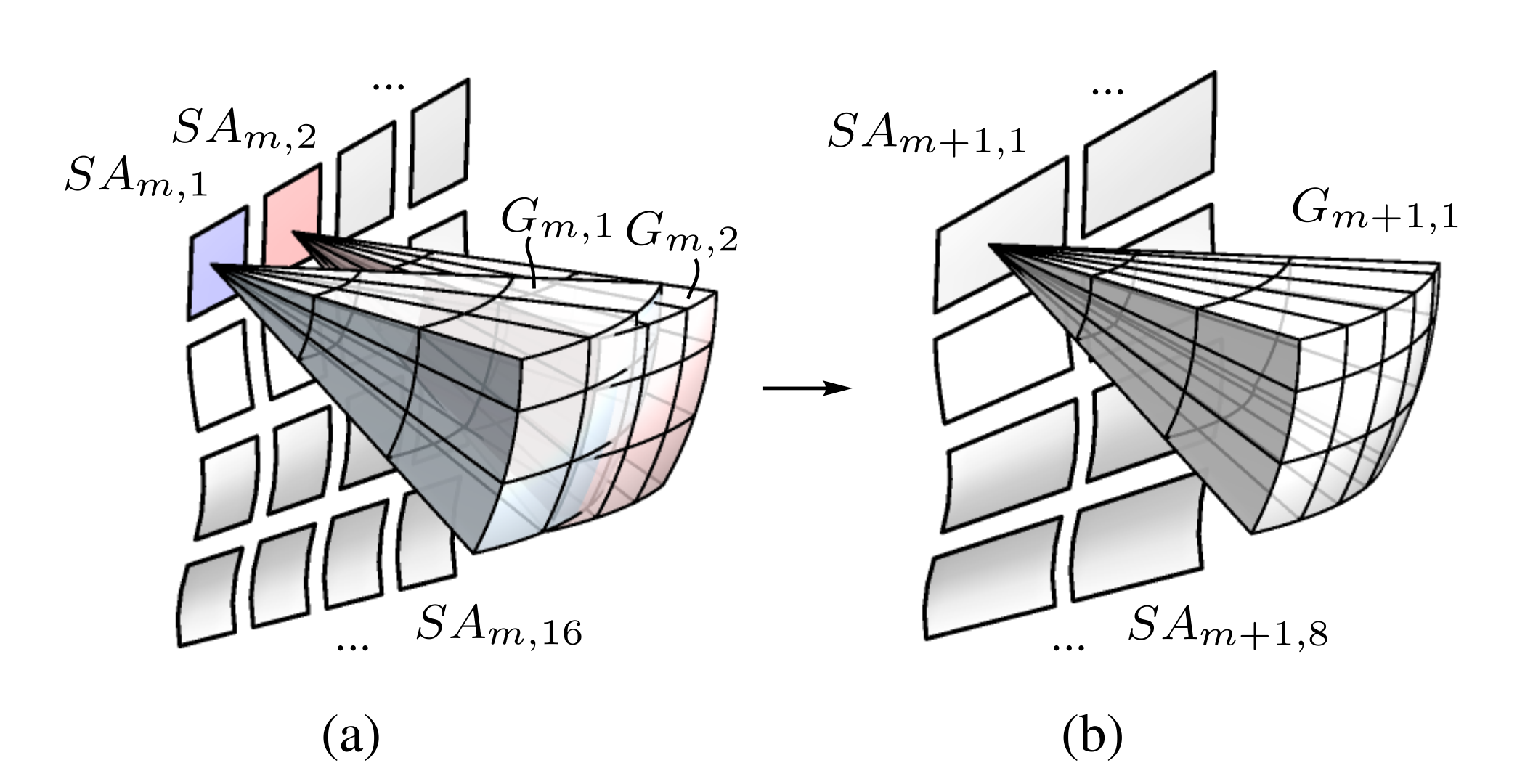}
\caption{Iteration step of the FFBP imaging process. The subarrays and the sampling grids of each subimage for the corresponding subarray are displayed. (a) Level $m$. (b) Level $m+1$.}
\label{fig_pcsffbp}
\end{figure}

The factorization technique as utilized in the fast factorized backprojection (FFBP) algorithm is introduced to effectively reduce the computational burden of BPA by employing a divide-and-conquer approach, where the reconstruction process is decomposed into multiple levels featuring smaller subarrays divided from the complete array and corresponding subimages downsampled via spectrum compression. FFBP algorithm iterates through the factorization levels to combine the subarrays, creating the next-level subimages, and the final imaging result as shown in Fig. \ref{fig_pcsffbp}.

In detail, the complete synthetic aperture is first decomposed into multiple smaller subarrays $SA_{1,n}$ at level 1, and the subimages are obtained through BPA, resulting in subimages $f_{1,n}(\textbf{p})$. The subimages exhibit lower resolution due to the reduced aperture size, as indicated by \eqref{eq_dxyz_kxyzminmax} and \eqref{eq_dxyz_xyzminmax}. This reduction enables downsampling of the subimages on sampling grids $G_{1,n}$, thereby reducing the total computational burden of BPA, which is formulated as

\begin{equation}\label{}
\begin{aligned}
f_{1,n}(\textbf{p}) = \iint\limits_{SA_{1,n}}s(\textbf{p}',\tau)|_{\tau=\frac{2||\textbf{p}-\textbf{p}'||}{c}}d\textbf{p}',\textbf{p}\in G_{1,n}.
\end{aligned}
\end{equation}

Subsequently, the subimages sampled on the downsampled grids are iteratively combined into next-level subimages with larger subarrays and denser sampling grids until the final image of the complete synthetic aperture is reconstructed:
\begin{equation}\label{}
\begin{aligned}
f_{m,n}(\textbf{p})=f_{m-1,2n-1}(\textbf{p})+f_{m-1,2n}(\textbf{p}),\ \textbf{p}\in G_{m,n}.
\end{aligned}
\end{equation}

The computational load of the FFBP algorithm is directly determined by the design of the sampling grids of each subimage. 
An effective design method of the sampling grids for near-range MIMO imaging systems has been proposed in \cite{song2024fast}. Numerical optimization techniques are employed to obtain the minimum volume cuboid bounding boxes of the local spectrum support at every position on the subimages.
These bounding boxes are then transformed through spatial downconversions (SDC) and local linear transformations (LLT) to be maximally fitted in the unit box at the wavenumber domain origin. SDCs and LLTs are implemented as downconversion function multiplications and coordinate transformations, respectively. These techniques realized effective spectrum compressions and enabled the uniform sampling of the subimages in the transformed coordinates without any local oversampling or undersampling.

However, this method requires extensive numerical pre-calculations to overcome the complexity of the spectrum support bounding box representations under near-range MIMO configurations and relies on the discrete Helmholtz-Hodge decomposition (HHD) to implement SDC and LLT operations. This is prohibitive for handheld SAR systems as the array configuration varies for each scan, so the pre-calculations can not be performed in advance and reused for subsequent imaging operations.

To solve this problem for handheld SAR systems, the approximate minimum volume bounding boxes of the local spectrum supports are derived analytically based on the discussions in Section \ref{subseclsp}. 

\begin{figure}[htbp]\centering
\includegraphics[width=3.5in]{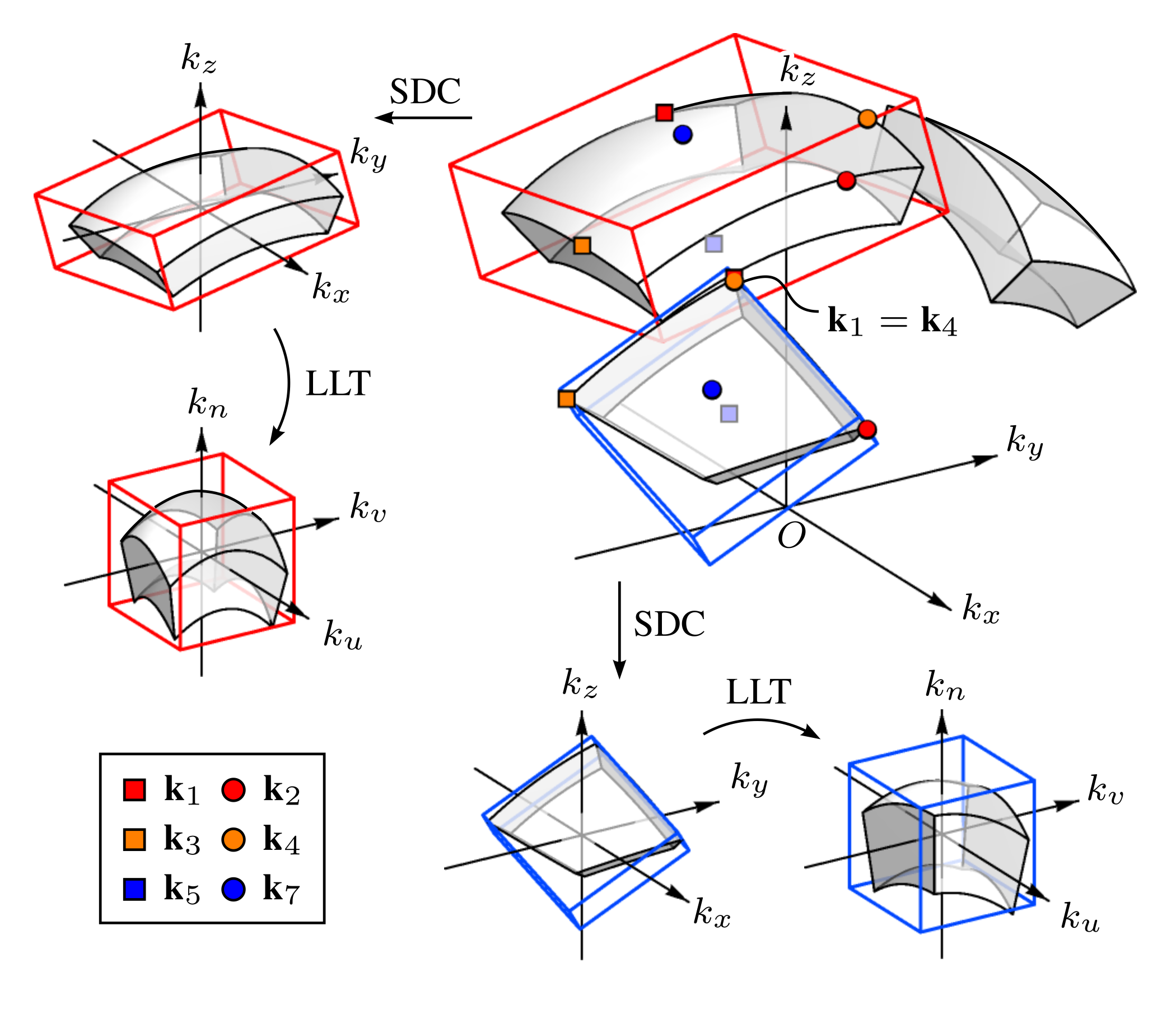}
\caption{Local spectrum supports of the subimages at different positions. The wavenumber domain key points $\textbf{k}_1$ to $\textbf{k}_7$ and the transformed spectrum supports after SDC and LLT at two arbitrary positions are also shown. }
\label{fig_locspcm}
\end{figure}

The local spectrum support of the subimages at arbitrary three positions in the imaging domain is illustrated in Fig. \ref{fig_locspcm}. The outlines of the local spectrum supports are sector-shaped regions with curved quadrilateral bases when projected towards the origin. To maximize the spectral occupancy of the subimages after the local spectrum transformations while still preserving the analytical representation, the approximate minimum volume bounding box of the local spectrum for subarray $SA$ and position $\textbf{p}$ in the imaging region can be expressed using key points in the wavenumber domain, which are defined as follows

\begin{equation}\label{}
\begin{aligned}
\textbf{k}_1=&\ \textbf{k}_0((x'_{\min,SA},y'_{0,SA},0),\textbf{p},k_{\max})\\
\textbf{k}_2=&\ \textbf{k}_0((x'_{\max,SA},y'_{0,SA},0),\textbf{p},k_{\max})\\
\textbf{k}_3=&\ \textbf{k}_0((x'_{0,SA},y'_{\min,SA},0),\textbf{p},k_{\max})\\
\textbf{k}_4=&\ \textbf{k}_0((x'_{0,SA},y'_{\max,SA},0),\textbf{p},k_{\max})\\
\textbf{k}_5=&(\textbf{k}_0((x'_{\min,SA},y'_{\min,SA},0),\textbf{p},k_{\min})\\    
    &+\textbf{k}_0((x'_{\min,SA},y'_{\max,SA},0),\textbf{p},k_{\min})\\
    &+\textbf{k}_0((x'_{\max,SA},y'_{\min,SA},0),\textbf{p},k_{\min})\\    
    &+\textbf{k}_0((x'_{\max,SA},y'_{\max,SA},0),\textbf{p},k_{\min}))/4\\
\textbf{k}_6=&\ 2k_{\max}\frac{\textbf{k}_5}{|\textbf{k}_5|},
\end{aligned}
\end{equation}
where $x'_{\min,SA}$, $x'_{\max,SA}$, $y'_{\min,SA}$, $y'_{\max,SA}$ denote the lower and upper bounds of $x'$ and $y'$ within the subarray $SA$, respectively. $x'_{0,SA}$, $y'_{0,SA}$ denote the nearest possible $x'$ and $y'$ within the subarray $SA$ to $x$ and $y$, which are formulated as
\begin{equation}\label{}
\begin{aligned}
x'_{0,SA}=&\begin{cases}
    x'_{\min,SA},&{\text{if}}\ x<x'_{\min,SA}\\ 
    x'_{\max,SA},&{\text{if}}\ x>x'_{\max,SA}\\
    x,&{\text{otherwise}} 
\end{cases}\\
y'_{0,SA}=&\begin{cases}
    y'_{\min,SA},&{\text{if}}\ y<y'_{\min,SA}\\ 
    y'_{\max,SA},&{\text{if}}\ y>y'_{\max,SA}\\
    y,&{\text{otherwise}}.
\end{cases}\\
\end{aligned}
\end{equation}

The wavenumber domain key point pairs $(\textbf{k}_1,\textbf{k}_2)$, $(\textbf{k}_3,\textbf{k}_4)$, and $(\textbf{k}_5,\textbf{k}_6)$ describe the span of the local spectrum support along the azimuth and range directions, respectively. As a result, the approximate minimum volume bounding box of the local spectrum support would be represented by a parallelepiped formed by the vectors $\textbf{k}_2-\textbf{k}_1$, $\textbf{k}_4-\textbf{k}_3$, and $\textbf{k}_6-\textbf{k}_5$. However, among the six key points, the curl of the vector field formed by $\textbf{k}_6$ is not zero with respect to $\textbf{p}$, which necessitates the complex HHD calculations during the SDC and LLT operations.

The integrating factor method is employed to transform $\textbf{k}_6$ to an irrotational field through space-variant scaling.
\begin{equation}\label{eq_spv_scl}
\begin{aligned}
\textbf{k}_7=\alpha'(x,y,z)\textbf{k}_6=\alpha(x,y,z)\textbf{k}_5
\end{aligned}
\end{equation}
\begin{equation}\label{eq_k7_curlfree}
\begin{aligned}
\nabla\times\textbf{k}_7=0,
\end{aligned}
\end{equation}
where $\alpha'(x,y,z)$ denotes the space-variant scaling factor and $\alpha(x,y,z)=2k_{\max}\alpha'(x,y,z)/|\textbf{k}_5|$. The solution is shown as follows
\begin{equation}\label{eq_spv_scl_comsol}
\begin{aligned}
\alpha(x,y,z) = \beta(r),
\end{aligned}
\end{equation}
where $\beta(\bullet)$ represents an arbitrary continuously differentiable function. $r$ denotes the sum of the distances between spatial point $\textbf{p}$ and the four vertices of the subarray, which is formulated as
\begin{equation}\label{eq_4r_sum}
\begin{aligned}
r=&\sqrt{(x-x'_{\min,SA})^2+(y-y'_{\min,SA})^2+z^2}\\
+&\sqrt{(x-x'_{\min,SA})^2+(y-y'_{\max,SA})^2+z^2}\\
+&\sqrt{(x-x'_{\max,SA})^2+(y-y'_{\min,SA})^2+z^2}\\
+&\sqrt{(x-x'_{\max,SA})^2+(y-y'_{\max,SA})^2+z^2}.
\end{aligned}
\end{equation}

Equations \eqref{eq_spv_scl}, \eqref{eq_spv_scl_comsol}, and \eqref{eq_4r_sum} describe the general solution for transforming $\textbf{k}_6$ to an irrotational vector field through the integrating factor method. The arbitrary function $\beta(\bullet)$ is then solved as follows so that the resulting $\textbf{k}_7$ remains the same as $\textbf{k}_6$ on the boresight axis of the subarray.
\begin{equation}\label{}
\begin{aligned}
\beta(r)=\frac{k_{\max}r}{k_{\min}\sqrt{r^2-4(x'_{\max}-x'_{\min})^2-4(y'_{\max}-y'_{\min})^2}}.
\end{aligned}
\end{equation}

The positions of the key points $\textbf{k}_1$, $\textbf{k}_2$, $\textbf{k}_3$, $\textbf{k}_4$, $\textbf{k}_5$, and $\textbf{k}_7$ for different positions $\textbf{p}$ are indicated in Fig. \ref{fig_locspcm}. Additionally, the parallelepipeds formed by the vectors $\textbf{k}_2-\textbf{k}_1$, $\textbf{k}_4-\textbf{k}_3$, and $\textbf{k}_7-\textbf{k}_5$ are illustrated with wireframes. Based on this result, SDC and LLT are subsequently employed to transform the local spectrum bounding box into the unity box at the wavenumber domain origin to enable the compressed uniform sampling in the transformed coordinates.

SDC is achieved by multiplying the subimage in the spatial domain with a space-variant phase term, which is
\begin{equation}\label{eq_sdc_mul}
\begin{aligned}
f'_{m,n}(\textbf{p})=f_{m,n}(\textbf{p})e^{-j\Phi},
\end{aligned}
\end{equation}
where the gradient of $\Phi$ should be the center of local spectrum support $\textbf{k}_c$, that is
\begin{equation}\label{eq_dp_kc}
\begin{aligned}
\nabla{\Phi}=\textbf{k}_c,
\end{aligned}
\end{equation}
and $\textbf{k}_c$ is expressed using the wavenumber domain key points
\begin{equation}\label{}
\begin{aligned}
\textbf{k}_c=\frac{\textbf{k}_5+\textbf{k}_6}{2}\approx\frac{\textbf{k}_5+\textbf{k}_7}{2}.
\end{aligned}
\end{equation}

The local spectrum supports will be shifted to the wavenumber domain origin after SDC as illustrated in Fig. \ref{fig_locspcm}, significantly reducing the required sampling rates of the subimage. 

\begin{figure*}[ht] \centering
\begin{equation}\label{eq_sdc_phi}
\begin{aligned}
\Phi=\frac{k_{\max}}{4}\sqrt{r^2-4(x'_{\max}-x'_{\min})^2-4(y'_{\max}-y'_{\min})^2}+\frac{k_{\min}}{4}r.
\end{aligned}
\end{equation}
\end{figure*}

Subsequently, LLT is utilized to transform the bounding box of the local spectrum support into a unit box at the origin of the wavenumber domain to further improve spectral occupancy. In contrast to \cite{song2024fast}, where only scaling and rotation transforms are employed, shearing transforms are also incorporated here as an extension. This is indicated by the fact that the estimated bounding boxes of the local spectrum support are no longer rectangular parallelepipeds.

The three generating vectors of the local spectrum support parallelepipeds are formulated as
\begin{equation}\label{}
\begin{aligned}
\textbf{v}_1&=\textbf{k}_2-\textbf{k}_1\\
\textbf{v}_2&=\textbf{k}_4-\textbf{k}_3\\
\textbf{v}_3&=\textbf{k}_7-\textbf{k}_5.
\end{aligned}
\end{equation}

The linear transforms, including scaling, rotation, and shearing transforms, in the wavenumber domain for the local spectrum support are incorporated into the linear coordinate transform matrix $\textbf{T}_k$. The LLT operation in the wavenumber domain can be formulated as
\begin{equation}\label{eq_llt_spcm_trans}
\begin{aligned}
2\pi\textbf{E}=\textbf{T}_k[\textbf{v}_1 \ \textbf{v}_2 \ \textbf{v}_3],
\end{aligned}
\end{equation}
where \textbf{E} is a unit matrix. Therefore, the spectrum of the subimage after SDC and LLT is confined within a $2\pi$ box centered at the origin, which corresponds to unity sampling rates along each dimension.

$\textbf{T}_k$ is accordingly solved as
\begin{equation}\label{}
\begin{aligned}
\textbf{T}_k=2\pi [\textbf{v}_1 \ \textbf{v}_2 \ \textbf{v}_3]^{-1}.
\end{aligned}
\end{equation}

According to the transformation properties of the Fourier transform, the required local linear coordinate transformation $\textbf{T}_s$ in the spatial domain is derived as
\begin{equation}\label{eq3_lltmatrix}
\begin{aligned}
\textbf{T}_s=(\textbf{T}_k^T)^{-1}=\frac{1}{2\pi}\begin{bmatrix}\textbf{v}_1^T\\\textbf{v}_2^T\\\textbf{v}_3^T\end{bmatrix}.
\end{aligned}
\end{equation}

To implement LLT on the subimages in the spatial domain, a coordinate transform $(x,y,z)\mapsto(u,v,n)=(u(x,y,z),v(x,y,z),n(x,y,z))$ which is locally linearized as the required local linear transformations is established:
\begin{equation}\label{eq_duvw_relp}
\begin{aligned}
\begin{bmatrix}
\frac{{\partial}u}{{\partial}x}&\frac{{\partial}u}{{\partial}y}&\frac{{\partial}u}{{\partial}z}\\
\frac{{\partial}v}{{\partial}x}&\frac{{\partial}v}{{\partial}y}&\frac{{\partial}v}{{\partial}z}\\
\frac{{\partial}n}{{\partial}x}&\frac{{\partial}n}{{\partial}y}&\frac{{\partial}n}{{\partial}z}
\end{bmatrix}
=\textbf{T}_s=
\frac{1}{2\pi}
\begin{bmatrix}
\textbf{v}_1^T\\
\textbf{v}_2^T\\
\textbf{v}_3^T
\end{bmatrix}.
\end{aligned}
\end{equation}

According to \eqref{eq_duvw_relp}, $u(x,y,z)$, $v(x,y,z)$, and $n(x,y,z)$ are the scalar potential functions of the vector fields ${\textbf{v}_1}/{(2\pi)}$, ${\textbf{v}_2}/{(2\pi)}$, and ${\textbf{v}_3}/{(2\pi)}$ respectively, and are solved as \eqref{eq_uvw_xyz}.

\begin{figure*}[ht] \centering
\begin{equation}\label{eq_uvw_xyz}
\begin{aligned}
u=&\ \frac{k_{\max}}{\pi}(\sqrt{(x-x'_{\min,SA})^2+(y-y'_{0,SA})^2+z^2}-\sqrt{(x-x'_{\max,SA})^2+(y-y'_{0,SA})^2+z^2})\\
v=&\ \frac{k_{\max}}{\pi}(\sqrt{(x-x'_{0,SA})^2+(y-y'_{\min,SA})^2+z^2}-\sqrt{(x-x'_{0,SA})^2+(y-y'_{\max,SA})^2+z^2})\\
n=&\ \frac{k_{\max}}{4\pi}\sqrt{r^2-4(x'_{\max}-x'_{\min})^2-4(y'_{\max}-y'_{\min})^2}-\frac{k_{\min}}{4\pi}r.
\end{aligned}
\end{equation}
\end{figure*}

The support range of the local spectra after SDC and LLT is illustrated in Fig. \ref{fig_locspcm}, exhibiting the well-confined spectrum distribution within the unit box at the origin of the wavenumber domain. In conclusion, the subimages of each subarray can now be expressed with a minimal number of sampling points by performing SDC (\ref{eq_sdc_mul}) and sampling on the uniform sampling grids under the transformed coordinates $(u,v,n)$ (\ref{eq_uvw_xyz}). 

\section{Implementation and Complexity Analysis}\label{sec_impcmp}

\subsection{Implementation}\label{subsec_imp}

\begin{figure*}[ht]\centering
\includegraphics[width=7in]{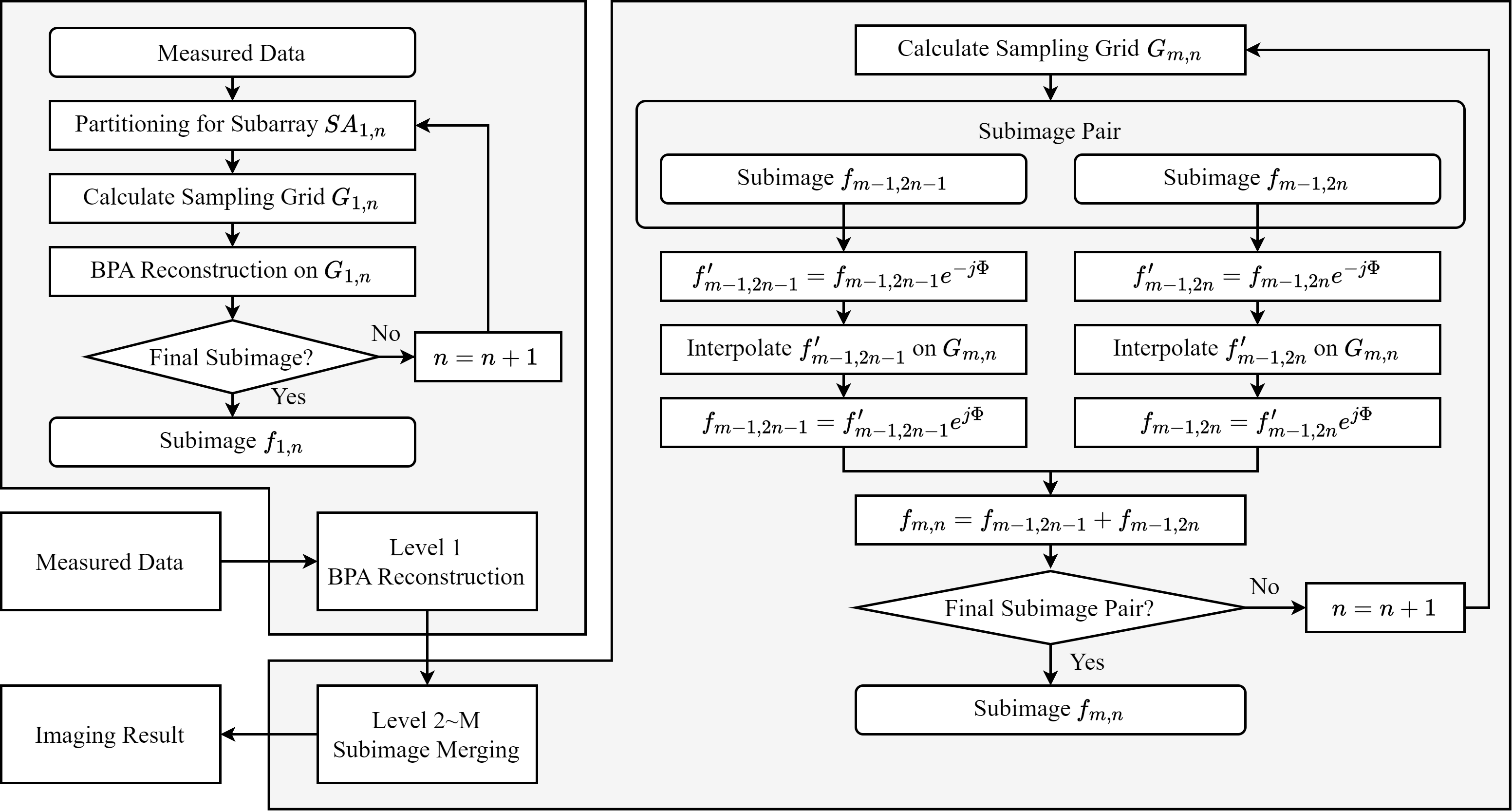}
\caption{Flowchart of the proposed algorithm.}
\label{fig_algimp}
\end{figure*}

With the analytical spectrum compression methods for handheld SAR subimages discussed above, the implementation of the proposed handheld SAR imaging algorithm (HHFFBPA) can now be developed. The flowchart is illustrated in Fig. \ref{fig_algimp}.

For a factorization level of $M$, the complete synthetic array is first decomposed into $2^{M-1}$ subarrays at the first level. Then, the uniform sampling grid with unity sampling rates in $(u,v,n)$ coordinates for each level 1 subarray is calculated and then converted to non-uniform sampling grids in $(x,y,z)$ coordinates through \eqref{eq_uvw_xyz}. BPA is subsequently employed to reconstruct the corresponding subimages for each subarray on the corresponding sampling grid. By applying SDC and LLT to exploit the local spectral properties, the required number of sampling points of the subimages and the computational burden of the BPA reconstructions are significantly reduced. 

The final image is obtained by coherently accumulating the subimages. This is achieved by multi-level interpolation operations on the adjacent subimages at each subsequent level. The $2^{M-1}$ subimages at level 1 form $2^{M-2}$ subimage pairs containing the subimage $f_{1,2n-1}$ and $f_{1,2n}$. To avoid aliasing during interpolations, the subimages are first converted to their compressed forms $f'_{1,2n-1}$ and $f'_{1,2n}$ through spatial downconversion described in \eqref{eq_sdc_mul} and \eqref{eq_sdc_phi}. These level 1 subimages with well-defined unity spectral support ranges are then interpolated onto the sampling grid $G_{2,n}$ at the next level. Then, the spatial downconversion is reverted by multiplying the subimages with the conjugate spatial downconversion functions to ensure a coherent accumulation of the subimages. Consequently, the next-level subimages sampled on the subimage grid $G_{2,n}$ are obtained. This process is repeated to obtain the $2^{M-m}$ subimages at level $m$ and the final imaging result $f_{M,1}$ at level M.

\subsection{Complexity}

The computational complexity of the proposed algorithm is discussed analytically. First, range compressions are performed on the measured signal according to \eqref{eq2_rangecompression}. The number of basic operations of the range compressions is
\begin{equation}\label{}
\begin{aligned}
N_{ops,RC} = C_1N_{A}N_f\log_2N_f,
\end{aligned}
\end{equation}
where $N_{A}$ is the number of elements of the complete synthetic array, $N_f$ is the number of frequency samples and $C_1$ is an implementation-dependent constant. 

The numbers of basic operations of the subsequent BPA reconstruction and subimage merging steps are
\begin{equation}\label{eq4_nopsbpa}
\begin{aligned}
N_{ops,BPA}=\sum\limits_{n=1}^{2^{M-1}}C_2N_{SA_{1,n}}N_{s,f_{1,n}}
\end{aligned}
\end{equation}
\begin{equation}\label{eq4_nopsinterp}
\begin{aligned}
N_{ops,Interp}=\sum\limits_{m=2}^{M}\sum\limits_{n=1}^{2^{M-m}}2C_3N_{s,f_{m,n}},
\end{aligned}
\end{equation}
where $N_{s,f_{m,n}}$ is the number of sampling points of the subimage $f_{m,n}$ after spectrum compressions through SDC and LLT, $N_{SA_{1,n}}$ is the number of array elements of the subarray $SA_{1,n}$ and $C_2$, $C_3$ are implementation-dependent constants. $N_{SA_{1,n}}$ satisfies
\begin{equation}\label{}
\begin{aligned}
\sum\limits_{n=1}^{2^{M-1}}N_{SA_{1,n}}=N_{A}.
\end{aligned}
\end{equation}

Therefore, the total number of operations of HHFFBPA is
\begin{equation}\label{}
\begin{aligned}
N_{ops}=N_{ops,RC}+N_{ops,BPA}+N_{ops,interp}.
\end{aligned}
\end{equation}

Considering that $N_{s,f_{m,n}}$ is higher for subarrays positioned near the center of the imaging region due to the higher spatial resolutions of the subimages according to \eqref{eq_dxyz_xyzminmax}. Therefore, the upper limit of $N_{s,f_{m,n}}$ is the number of sampling points of $f_{m,0}$ for the subarray $SA_{m,0}$ located near the center of the imaging region.
\begin{equation}\label{eq4_subimgsmppresimp}
\begin{aligned}
N_{s,f_{m,n}} \leq N_{s,f_{m,0}} = \iiint\limits_{D} |\det T_{s,SA_{m,0}}|dxdydz,
\end{aligned}
\end{equation}
where $T_{s,SA_{m,0}}$ is the local transformation matrix for subarray $SA_{m,0}$ described in \eqref{eq3_lltmatrix}. 

However, $T_{s,SA_{m,0}}$ is space-variant, and the precise result of $N_{s,f_{m,0}}$ is hard to analytically derive. For simplicity, the determinant of the transformation matrix for the center of the imaging region, $|\det T_{s0,SA_{m,0}}|$, is taken as an approximation of the mean value of the determinant as implicitly referenced in the integral. Then, \eqref{eq4_subimgsmppresimp} is simplified as
\begin{equation}\label{}
\begin{aligned}
N_{s,f_{m,0}}
\approx\ &V(D)|\det T_{s0,SA_{m,0}}|\\
=\ &\frac{V(D)L^2k_{\max}^2(k_{\max}-\frac{2k_{\min}z_{\text{mid}}}{\sqrt{2^{m-M+1}L^2+4z_{\text{mid}}^2}})}{2^{M-m-2}\pi^3(2^{m-M}L^2+4z_{\text{mid}}^2)},
\end{aligned}
\end{equation}
where $z_{\text{mid}}$ is the distance between the center of the synthetic array and the center of the imaging region, and $L$ is the size of the complete synthetic array.

The ratio of $N_{s,f_{m+1,0}}$ and $N_{s,f_{m,0}}$ is derived to analyze the asymptotic relationship of the sampling points of subimages between the adjacent factorization levels.
\begin{equation}\label{eq4_smpratiopresimp}
\begin{aligned}
\frac{N_{s,f_{m+1,0}}}{N_{s,f_{m,0}}}
=\ &2\frac{L^2 2^{m-M}+4 z_{\text{mid}}^2}{L^2 2^{m-M+1}+4 z_{\text{mid}}^2}\\
&\times \frac{k_{\max }-\frac{2 k_{\min } z_{\text{mid}}}{\sqrt{L^2 2^{m-M+2}+4 z_{\text{mid}}^2}}}{k_{\max }-\frac{2k_{\min } z_{\text{mid}}}{\sqrt{L^2 2^{m-M+1}+4 z_{\text{mid}}^2}}}.
\end{aligned}
\end{equation}

For handheld SAR systems, the array aperture size $L$ is of the same order as $z_{\text{mid}}$. Therefore, the subarray aperture is greatly smaller than $z_{\text{mid}}$ and $L^2 2^{m-M+1} \ll 4 z_{\text{mid}}^2$. \eqref{eq4_smpratiopresimp} is further simplified as
\begin{equation}\label{eq_smpratiosim2}
\begin{aligned}
\frac{N_{s,f_{m+1,0}}}{N_{s,f_{m,0}}}\approx 2.
\end{aligned}
\end{equation}

Please note that while the above approximation is no longer optimal for the final few factorization levels where the apertures of the subarrays are no longer small enough, they contribute a diminishing percentage to the total computational costs of the interpolation operations as the problem size $N$ grows. Applying \eqref{eq_smpratiosim2} to \eqref{eq4_nopsbpa} and \eqref{eq4_nopsinterp}, the number of basic operations can now be further simplified
\begin{equation}\label{}
\begin{aligned}
N_{ops,BPA}
\approx\ &\sum\limits_{n=1}^{2^{M-1}}C_2\frac{N_{A}}{2^{M-1}}\frac{N_{s,f_{M,1}}}{2^{M-1}}\\
=\ &C_2\frac{N_{A}N_{s,f_{M,1}}}{2^{M-1}},
\end{aligned}
\end{equation}
\begin{equation}\label{}
\begin{aligned}
N_{ops,Interp}
\approx\ &\sum\limits_{m=2}^{M}\sum\limits_{n=1}^{2^{M-m}}2C_3\frac{N_{s,f_{M,1}}}{2^{M-m}}\\
=\ & 2C_3(M-1)N_{s,f_{M,1}}.
\end{aligned}
\end{equation}

As a result, the total number of operations is 
\begin{equation}\label{eq4_nopstotal}
\begin{aligned}
N_{ops}
\approx\ &C_1N_{A}N_f\log_2N_f+C_2\frac{N_{A}N_{s,f_{M,1}}}{2^{M-1}}\\
&+2C_3(M-1)N_{s,f_{M,1}}.
\end{aligned}
\end{equation}

From \eqref{eq4_nopstotal}, it is obvious that $N_{ops,BPA}$ is lower and $N_{ops,Interp}$ is higher if a higher number of factorization level $M$ is chosen. Additionally, a higher $M$ means more interpolation operations and therefore more interpolation errors. In practice, $M$ can be determined by trading imaging quality off computational cost.

The final computational complexity of the proposed HHFFBPA can now be determined. Let the problem size $N$ be defined as the order of the synthetic aperture dimension, that is $N_{A}\sim N^2$, and the number of frequency samples is of the same order of $N$, so $N_f\sim N$. Then $N_{ops}$ achieves its minimum when $M$ is of order $2\log{N}$ and $N_{ops}\sim N^3\log{N}$. For comparison, the computational complexity of BPA and EPC-RMA is also listed in Table \ref{tab_algocmpx}. In conclusion, HHFFBPA exhibits high computational efficiency with the same computational complexity as that of EPC-RMA.

\begin{table}[htbp]\centering
\caption{Computational Complexity of Different Imaging Algorithms}
\begin{tabular}{
>{\raggedright\arraybackslash}p{2.5cm}
>{\raggedright\arraybackslash}p{2.5cm}}
\toprule
Algorithm & Complexity\\
\midrule
BPA     & $\mathcal{O}(N^5)$ \\
EPC-RMA & $\mathcal{O}(N^3\log N)$ \\
HHFFBPA & $\mathcal{O}(N^3\log N)$ \\
\bottomrule
\end{tabular}
\label{tab_algocmpx}
\end{table}

\section{Numerical Simulations and Experiments} \label{sec_simexp}

\subsection{Simulation Configuration}

\begin{figure}[!htbp]\centering
\includegraphics[width=3.5in]{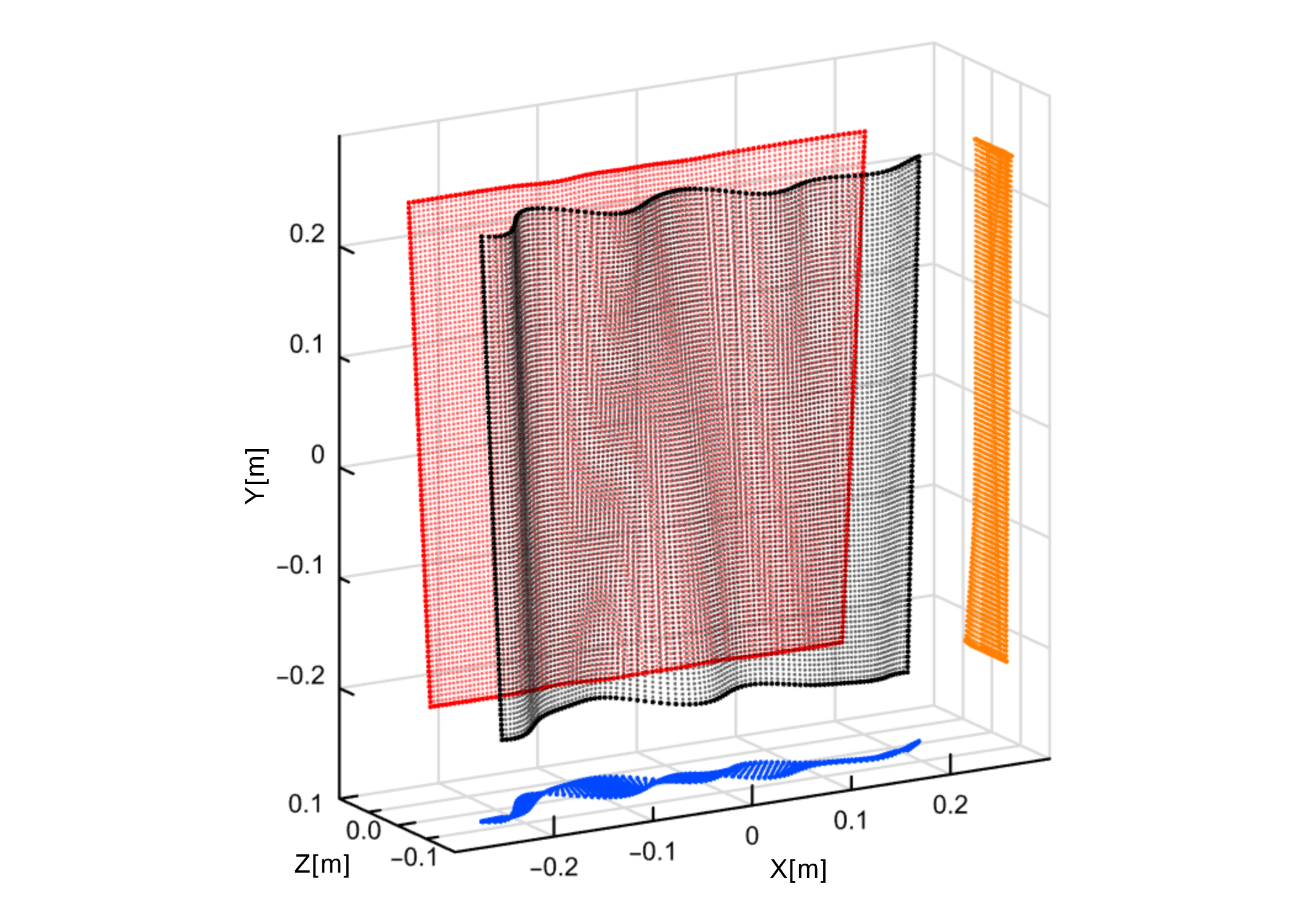}
\caption{Synthetic array element positions used in the numerical simulations and experiments. The projections of the elements on the corresponding planes are also shown.}
\label{fig_arrpos}
\end{figure}

The image quality and computational efficiency of the proposed algorithm are first verified through numerical simulations. Fig. \ref{fig_arrpos} illustrates the synthetic array element positions used in the numerical simulations and experiments. A linear array is scanned along the $x$-axis while incorporating position and angle fluctuations to simulate the trajectory uncertainties in handheld applications. The numbers of array elements and scanning positions are both 101 samples, forming an approximately \SI{0.45}{m}-by-\SI{0.45}{m} synthetic aperture. The depth fluctuations of the array elements are within \SI{8}{cm}. 

The working frequency in numerical simulation ranges from \SI{12}{\giga\hertz} to \SI{15}{\giga\hertz} with 24 equidistant frequency samples. The dimensions of the imaging region are \SI{0.5}{\meter}$\times$\SI{0.5}{\meter}$\times$\SI{0.5}{\meter}. The depth of the center of the imaging region is \SI{0.4}{\meter} away from the center of the synthetic aperture. The resolution of the final image in pixels is 101$\times$101$\times$51 which is derived from the spatial resolution limits of the radar system. 

The proposed algorithm is compared with the conventional algorithms, namely BPA and EPC-RMA. For EPC-RMA, phase corrections with the center of the imaging region as the reference position will be employed to mitigate the error introduced by approximating the deformed array with a planar array. However, there will still be significant defocusing and distortion at positions away from the center of the imaging region. 

All algorithms are implemented in C++ code and executed on a desktop computer with a 4.5-GHz AMD Ryzen 9 7950X central processing unit and 64-GB DDR5 random access memory, without utilizing parallelization or any other acceleration techniques.

\subsection{Simulation \Rmnum{1}: Point Scatterers}

\begin{figure}[!htbp]\centering
\includegraphics[width=8.6cm]{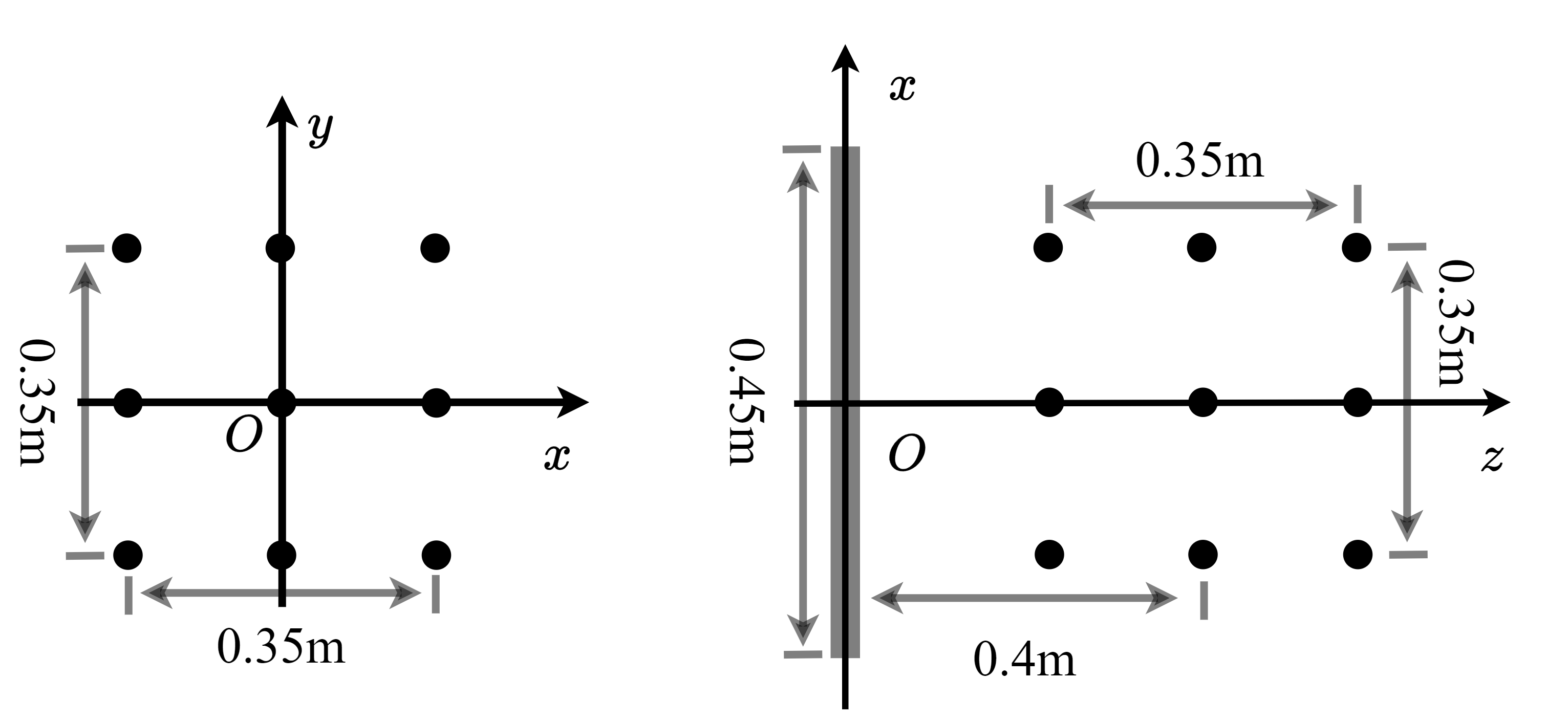}
\caption{Point scatterer configuration used in numerical simulations.}
\label{fig_simpnt}
\end{figure}

The imaging quality of different algorithms is qualitatively and quantitatively analyzed by conducting numerical simulations with point targets, as depicted in Fig. \ref{fig_simpnt}. The simulation scene consists of a 3{$\times$}3{$\times$}3 array of point scatterers with a spacing of \SI{0.175}{m} between the adjacent scatterers. The distance between the center scatterer and the center of the synthetic aperture is \SI{0.4}{m}.

\begin{figure}[!htbp]\centering
\begin{minipage}[t]{4.1cm}\centering
\includegraphics[width=4.1cm]{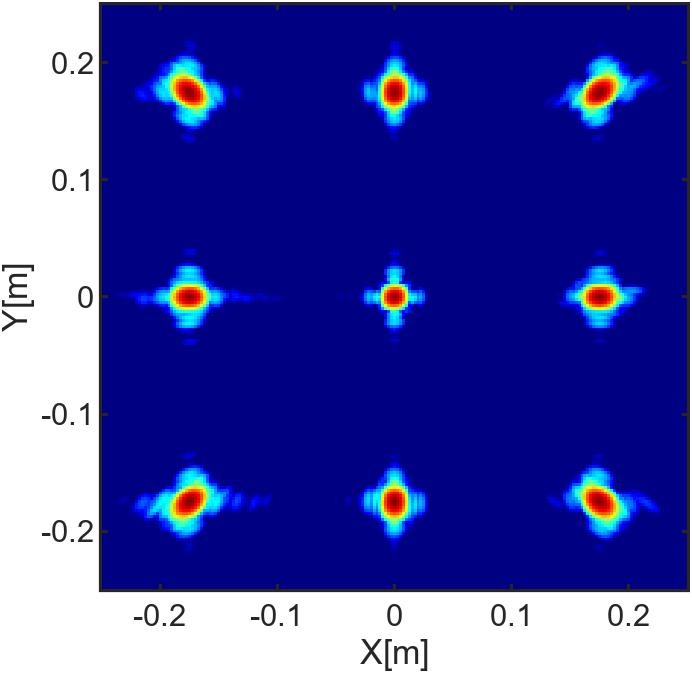}\\\footnotesize{(a)}
\end{minipage}
\begin{minipage}[t]{4.5cm}\centering
\includegraphics[width=4.5cm]{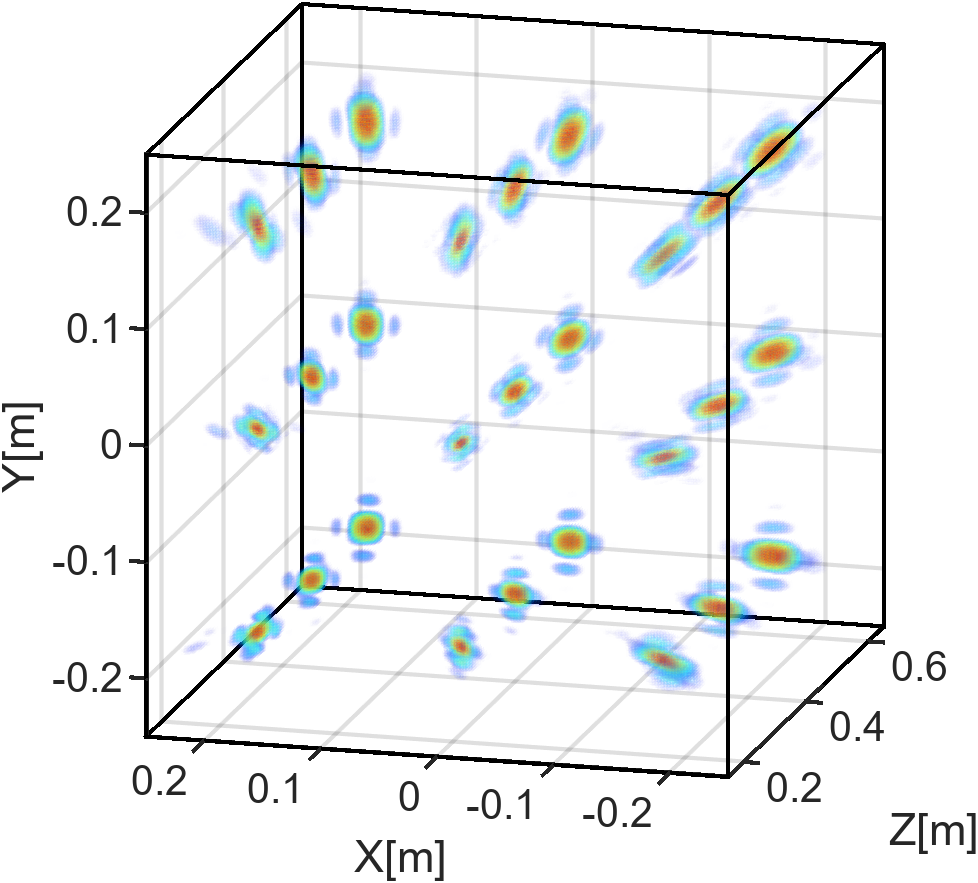}\\\footnotesize{(b)}
\end{minipage}
\begin{minipage}[t]{4.1cm}\centering
\includegraphics[width=4.1cm]{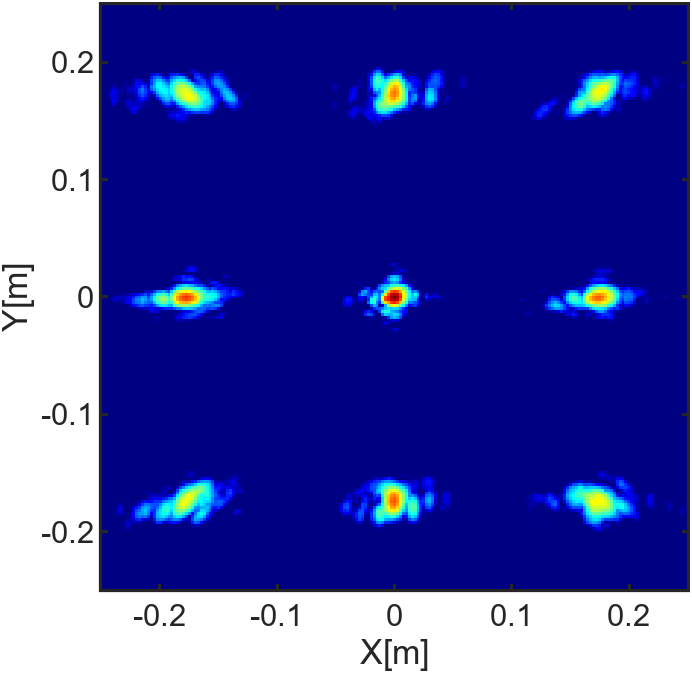}\\\footnotesize{(c)}
\end{minipage}
\begin{minipage}[t]{4.5cm}\centering
\includegraphics[width=4.5cm]{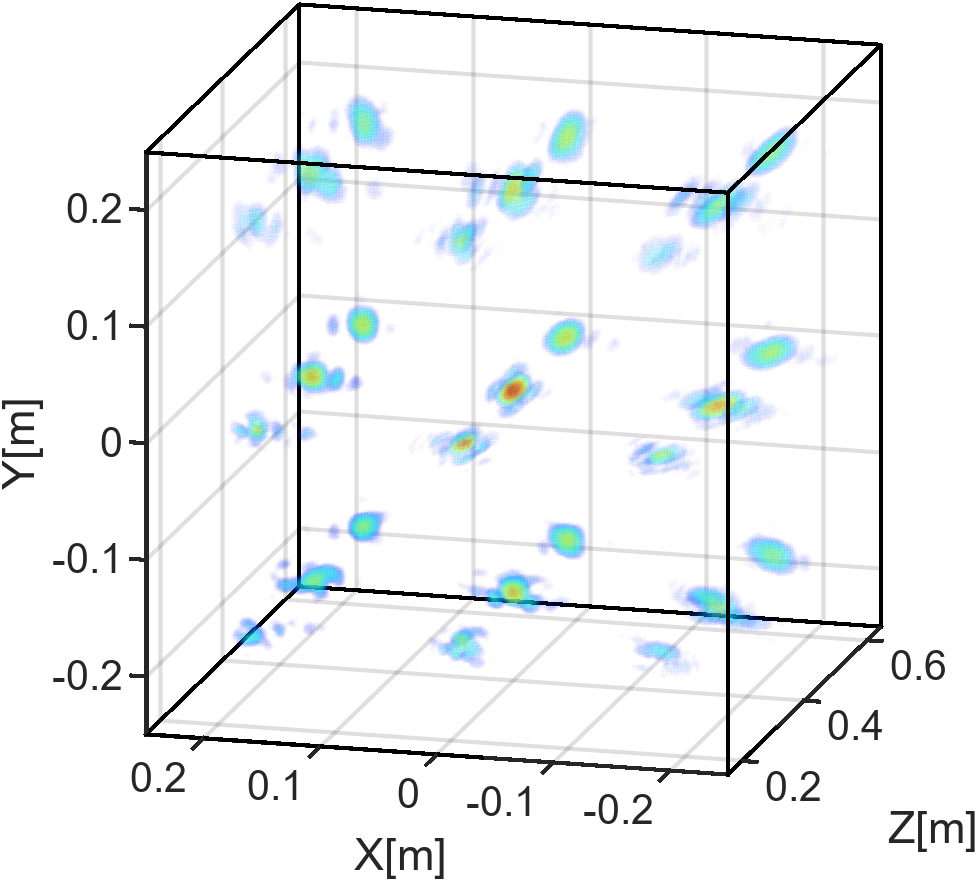}\\\footnotesize{(d)}
\end{minipage}
\begin{minipage}[t]{4.1cm}\centering
\includegraphics[width=4.1cm]{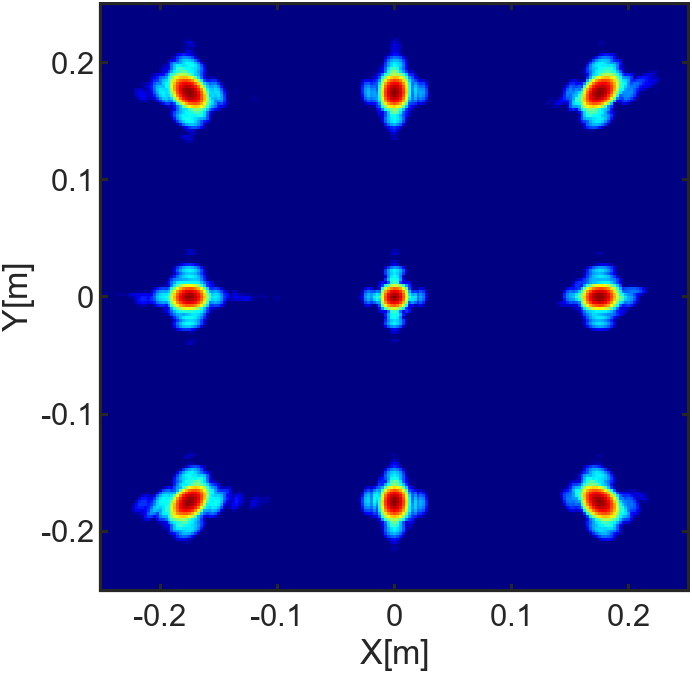}\\\footnotesize{(e)}
\end{minipage}
\begin{minipage}[t]{4.5cm}\centering
\includegraphics[width=4.5cm]{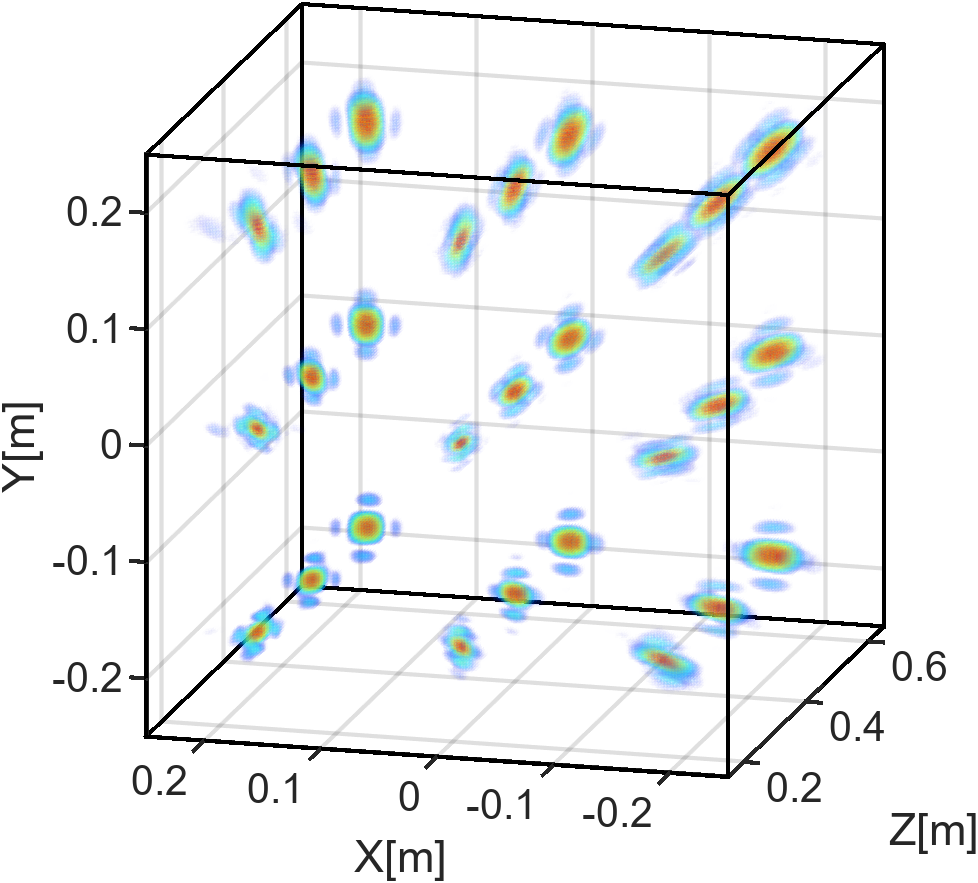}\\\footnotesize{(f)}
\end{minipage}
\begin{minipage}[t]{8.6cm}\centering
\includegraphics[width=6.0cm]{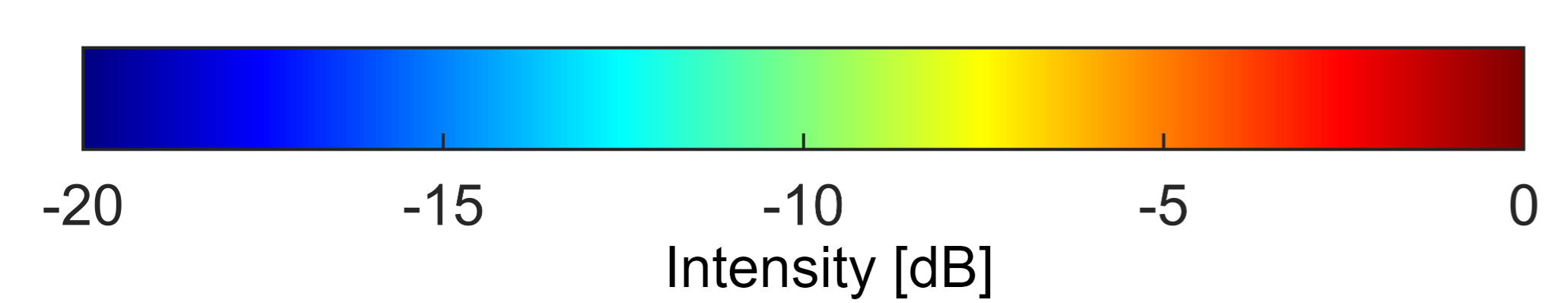}
\end{minipage} 
\caption{Imaging results of the different algorithms in Simulation \Rmnum{1}. (a)(c)(e) Maximum intensity projection results. (b)(d)(f) 3-D volume rendering results. (a)(b) Results of BPA. (c)(d) Results of EPC-RMA. (e)(f) Results of the proposed HHFFBPA.}
\label{fig_pntsimimg}
\end{figure}

The reconstruction results of the point scatterer scene with different algorithms are presented in Fig. \ref{fig_pntsimimg}. It is evident that the proposed algorithm HHFFBPA demonstrates comparable imaging quality to BPA, both in the central region and the squint region. Conversely, while EPC-RMA exhibits satisfactory focusing performance in the central region, it exhibits noticeable defocusing artifacts in the squint region, thereby significantly compromising the overall imaging quality.

\begin{figure}[!htbp]\centering
\begin{minipage}[t]{3.5in}\centering
\includegraphics[width=3.5in]{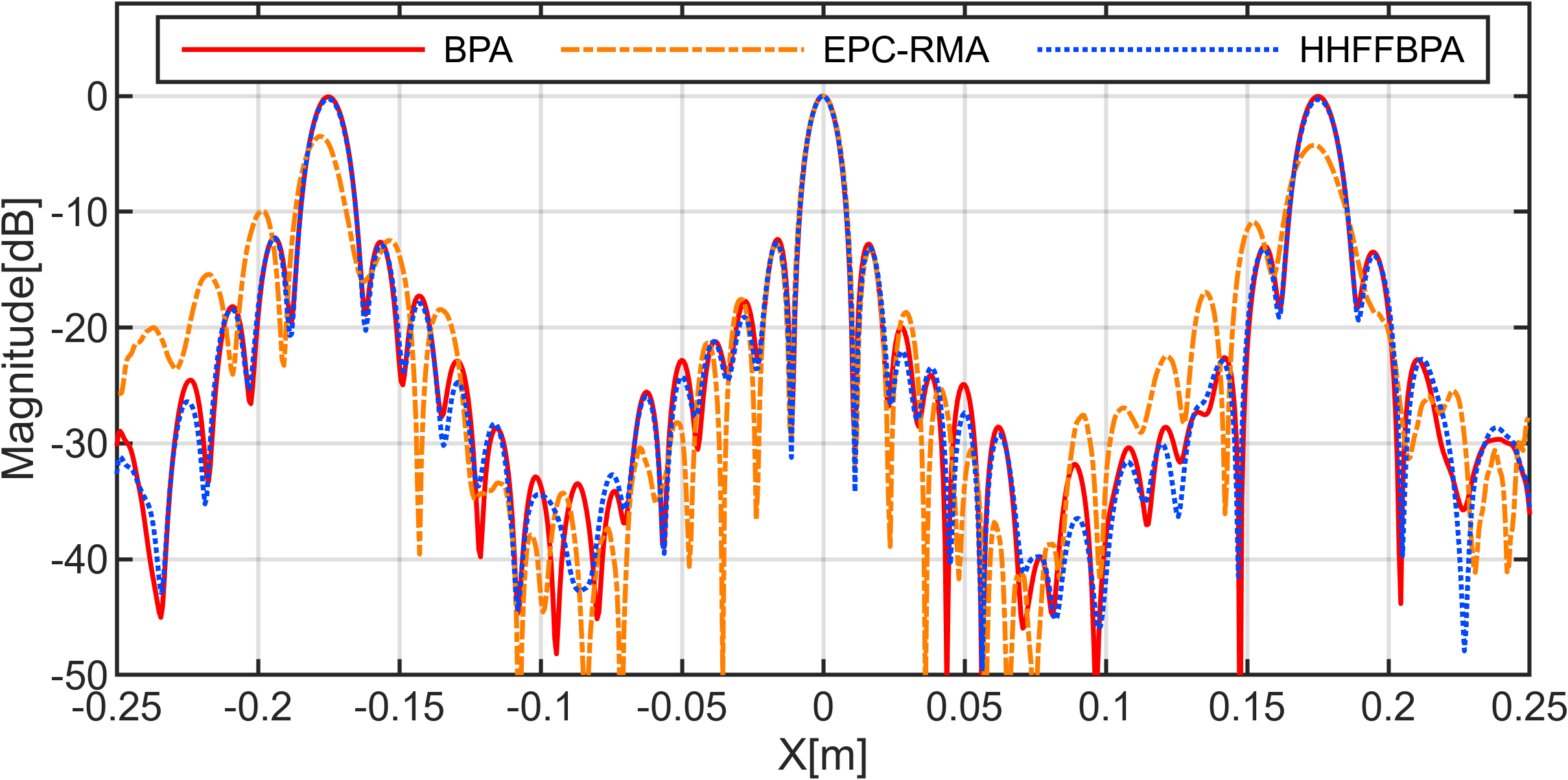}\\\footnotesize{(a)}
\end{minipage}
\begin{minipage}[t]{3.5in}\centering
\includegraphics[width=3.5in]{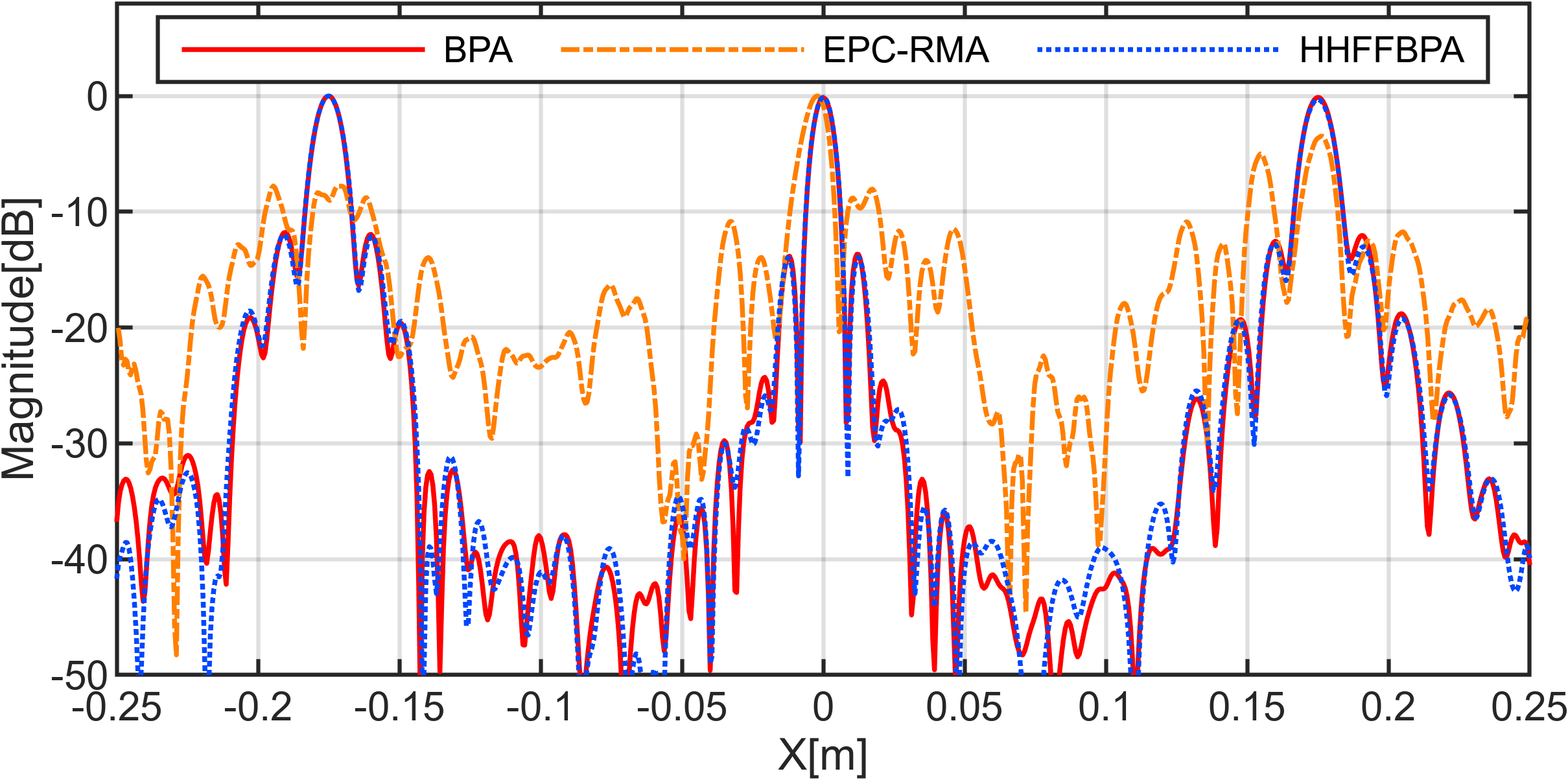}\\\footnotesize{(b)}
\end{minipage}
\caption{Azimuth slices of the imaging results of the different algorithms in Simulation \Rmnum{1}. (a) Slice at the center position. (b) Slice at the squint position.}
\label{fig_pntsimpsf}
\end{figure}

To visually demonstrate the focusing performance of different algorithms at various positions, Fig. \ref{fig_pntsimpsf} illustrates the slices of the results in Fig. \ref{fig_pntsimimg} along the $x$-axis at the center position ($x$, \SI{0}{m}, \SI{0.4}{m}) and the squint position ($x$, \SI{-0.175}{m}, \SI{0.225}{m}). The point spread functions (PSFs) of the proposed algorithm HHFFBPA exhibit remarkable consistency with BPA, which is also observed for EPC-RMA at the center position, indicating excellent focusing performance. However, for the squint targets, there is a significant deterioration in imaging quality for EPC-RMA compared to BPA and HHFFBPA. This discrepancy arises because of the increasing phase errors introduced by the planar array corrections, which only work perfectly at the reference position.

Quantitative metrics, including the mainlobe widths, peak sidelobe ratios (PSLR), and integral sidelobe ratios (ISLR), of the point spread functions at the center position (\SI{0}{m}, \SI{0}{m}, \SI{0.4}{m}) and the squint position (\SI{-0.175}{m}, \SI{0}{m}, \SI{0.4}{m}) in Fig. \ref{fig_pntsimpsf}(a) are listed in Table \ref{tab_pntsimpsfidx}. Consistent with the above analysis, the proposed algorithm demonstrates comparable focusing and sidelobe performance to BPA, with improvements of \SI{0.59}{mm}, \SI{5.61}{dB}, and \SI{5.54}{dB} over EPC-RMA at the squint position for mainlobe width, PSLR, and ISLR, respectively. Notably, these differences become more pronounced as the squint angle increases, as depicted by the profiles in Fig. \ref{fig_pntsimpsf}(b). Furthermore, Table \ref{tab_pntsimpsfidx} also presents the peak signal-to-noise ratio (PSNR) of the imaging results in Fig. \ref{fig_pntsimimg} compared to BPA.

\begin{table*}[htbp] \centering
\caption{Quantitative Metrics of the Imaging Results in Simulation \Rmnum{1}}
\begin{tabular}{
>{\raggedright\arraybackslash}m{1.5cm}
>{\raggedright\arraybackslash}m{1.5cm}
>{\raggedright\arraybackslash}m{1.5cm}
>{\raggedright\arraybackslash}m{1.5cm}
>{\raggedright\arraybackslash}m{1.5cm}
>{\raggedright\arraybackslash}m{1.5cm}
>{\raggedright\arraybackslash}m{1.5cm}
>{\raggedright\arraybackslash}m{1.5cm}}
\toprule
\multirow{2}{*}{Algorithms} & \multicolumn{3}{c}{Center Scatterer}  & \multicolumn{3}{c}{Squint Scatterer} & \multirow{2}{*}{PSNR {[}dB{]}} \\ 
\cline{2-7}
 & Mainlobe Width {[}mm{]} & PSLR {[}dB{]} & ISLR {[}dB{]} & Mainlobe Width {[}mm{]} & PSLR {[}dB{]} & ISLR {[}dB{]} &  \\ 
\midrule
BPA     & 9.68  & -12.77    & -9.91     & 11.55 & -12.28    & -9.90 & ----- \\
EPC-RMA & 10.05 & -13.07    & -10.96    & 12.29 & -6.54     & -4.45 & 25.46 \\
HHFFBPA & 9.68  & -13.04    & -10.63    & 11.70 & -12.15    & -9.99 & 45.98 \\ 
\bottomrule
\end{tabular}
\label{tab_pntsimpsfidx}
\end{table*}

The computation time of different algorithms is presented in Table \ref{tab_simalgotime}. The proposed algorithm HHFFBPA exhibits exceptional computational efficiency, being 62.09 times faster than BPA, due to the utilization of the factorization technique. However, it is still 10.79 times slower than EPC-RMA due to the efficient implementation of fast Fourier transforms and the lower number of numerical operations despite the same computational complexity. Considering both imaging quality and computational efficiency, the proposed algorithm HHFFBPA demonstrates remarkable imaging performance for handheld SAR systems.

\begin{table}[htbp]\centering
\caption{Computation Time of the Different Algorithms in Numerical Simulations}
\begin{tabular}{
>{\raggedright\arraybackslash}p{2.5cm}
>{\raggedright\arraybackslash}p{2.5cm}}
\toprule
Algorithms  & Computation time [s] \\
\midrule
BPA         & 187.52 \\
EPC-RMA     & 0.28 \\
HHFFBPA      & 3.02 \\
\bottomrule
\end{tabular}
\label{tab_simalgotime}
\end{table}

\subsection{Simulation \Rmnum{2}: Extended Target}

\begin{figure}[!htbp]\centering
\includegraphics[width=8.6cm]{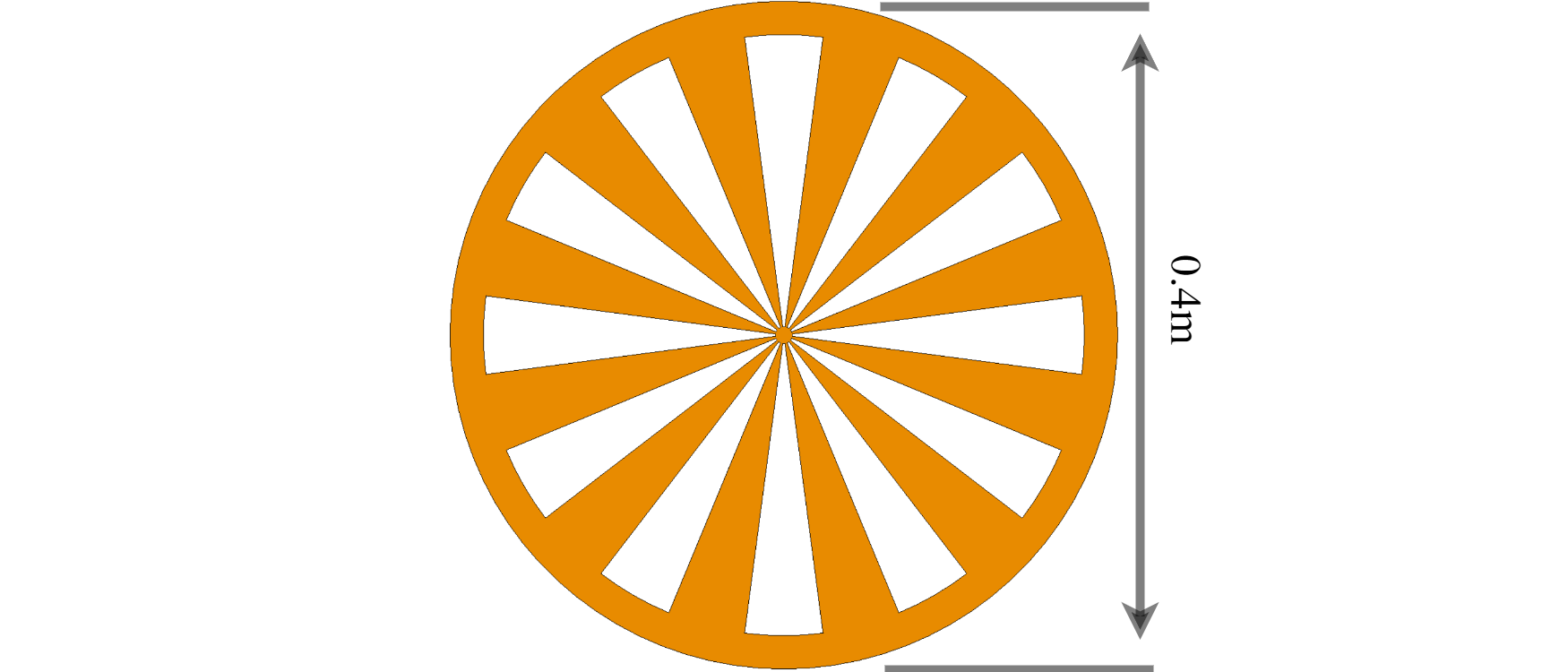}
\caption{Extended target used in Simulation \Rmnum{2}.}
\label{fig_simsms}
\end{figure}

The imaging performance of the proposed algorithm is further verified through numerical simulations with extended targets. A Siemens star with a diameter of \SI{0.4}{m}, depicted in Fig. \ref{fig_simsms}, is used in this simulation. The target is positioned at a distance of \SI{0.4}{m} from the center of the array. The other system parameters are kept consistent with the above point target simulation. The scatter data of the extended target is generated using the electromagnetic simulation tool FEKO with physical optics methods.

\begin{figure}[!htbp]\centering
\begin{minipage}[t]{4.1cm}\centering
\includegraphics[width=4.1cm]{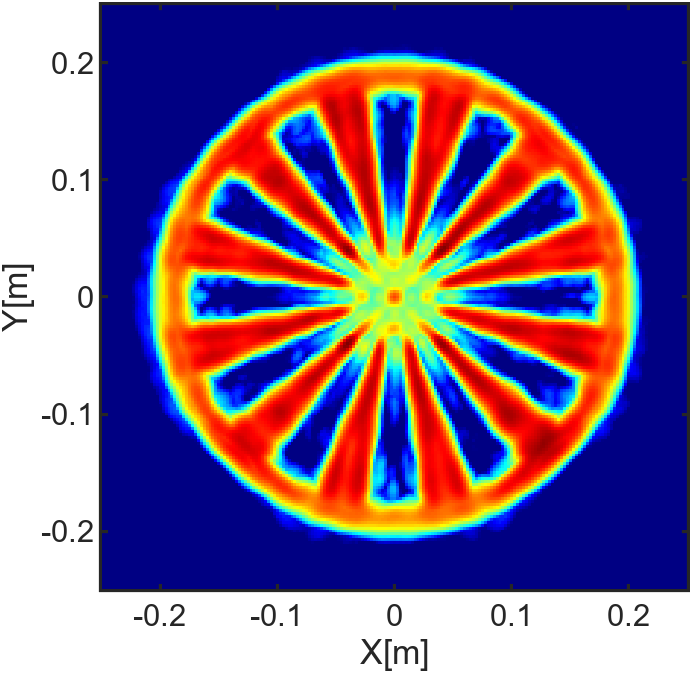}\\\footnotesize{(a)}
\end{minipage}
\begin{minipage}[t]{4.5cm}\centering
\includegraphics[width=4.5cm]{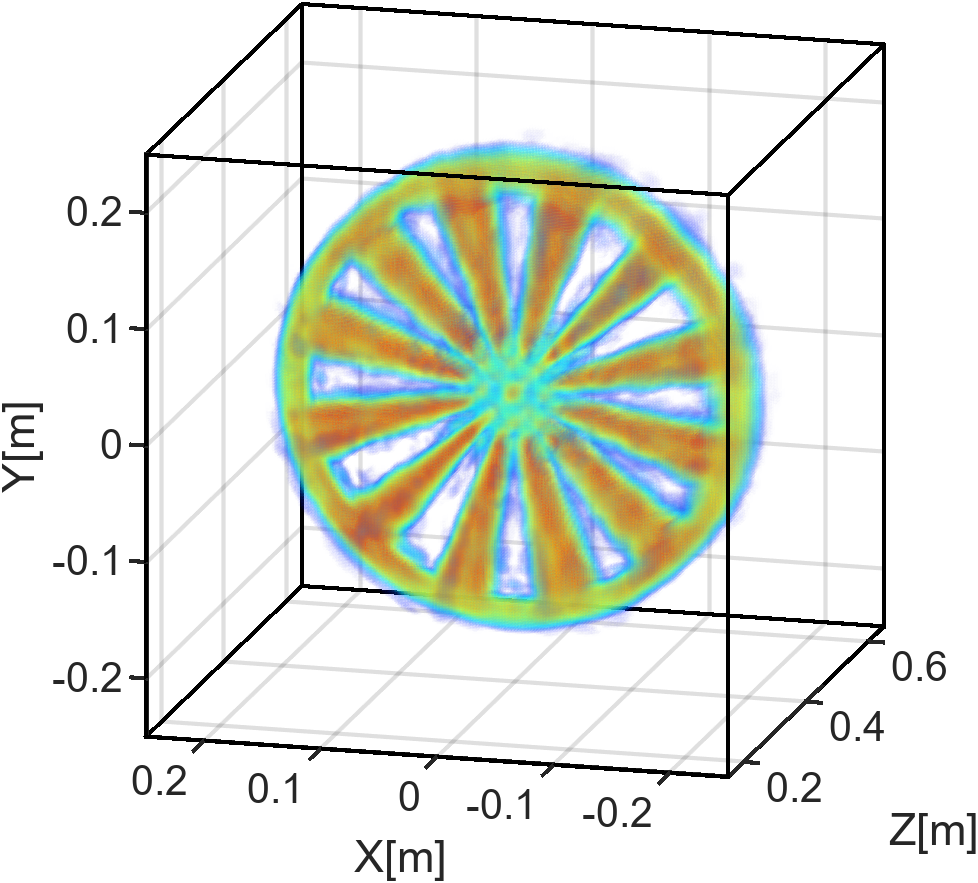}\\\footnotesize{(b)}
\end{minipage}
\begin{minipage}[t]{4.1cm}\centering
\includegraphics[width=4.1cm]{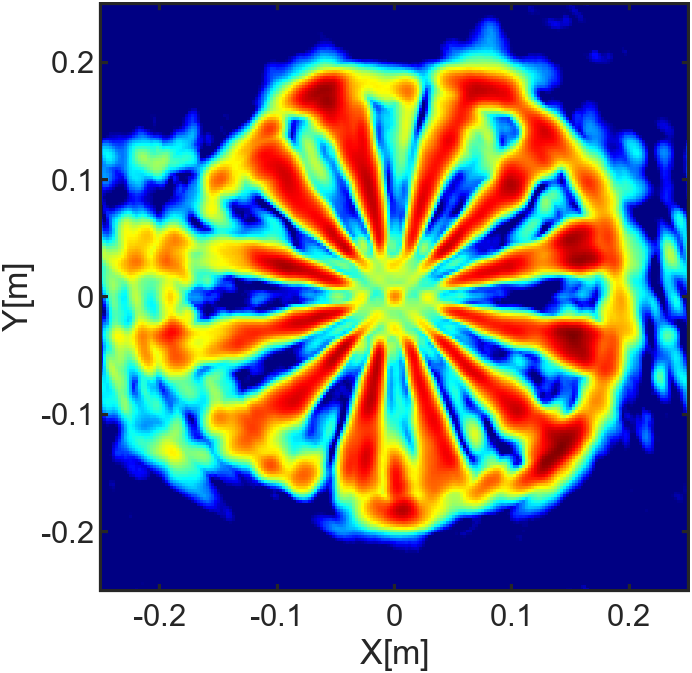}\\\footnotesize{(c)}
\end{minipage}
\begin{minipage}[t]{4.5cm}\centering
\includegraphics[width=4.5cm]{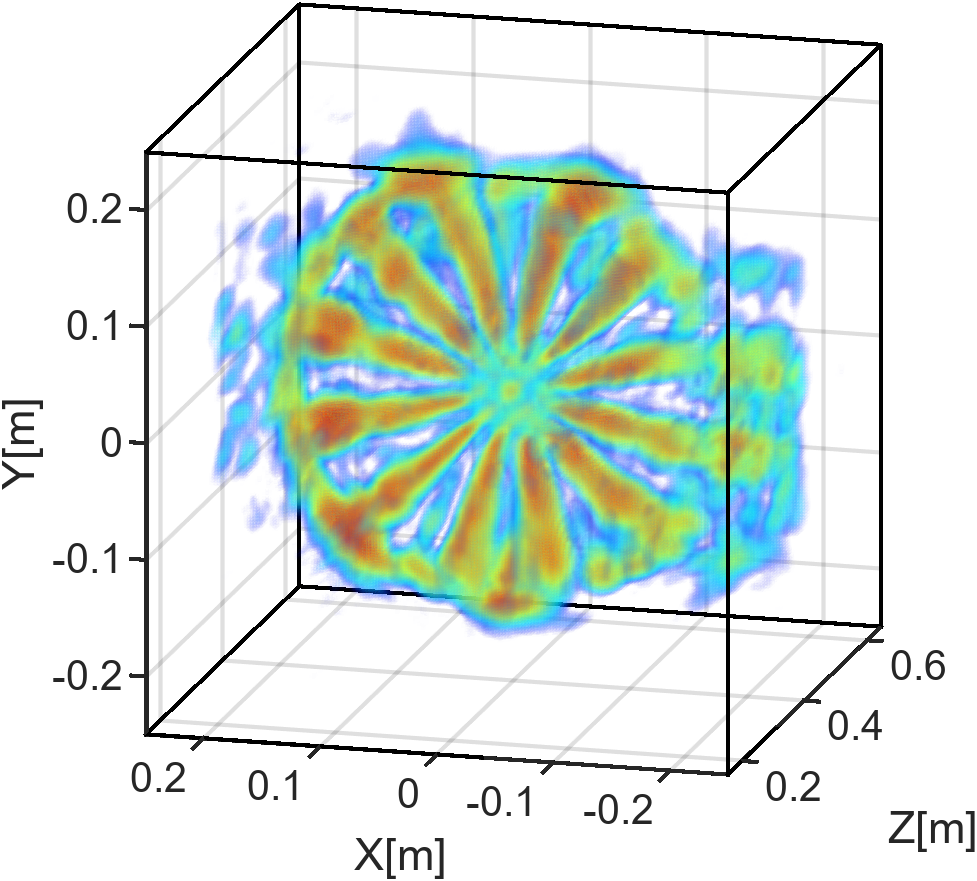}\\\footnotesize{(d)}
\end{minipage}
\begin{minipage}[t]{4.1cm}\centering
\includegraphics[width=4.1cm]{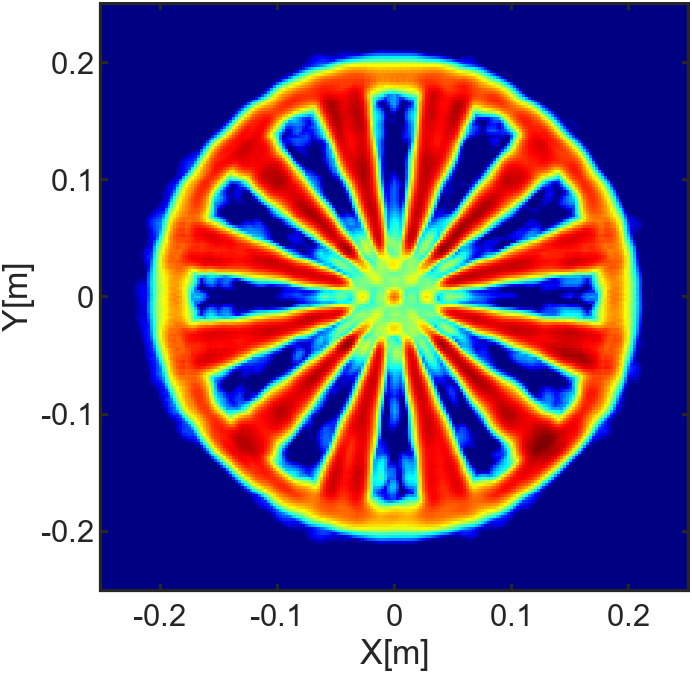}\\\footnotesize{(e)}
\end{minipage}
\begin{minipage}[t]{4.5cm}\centering
\includegraphics[width=4.5cm]{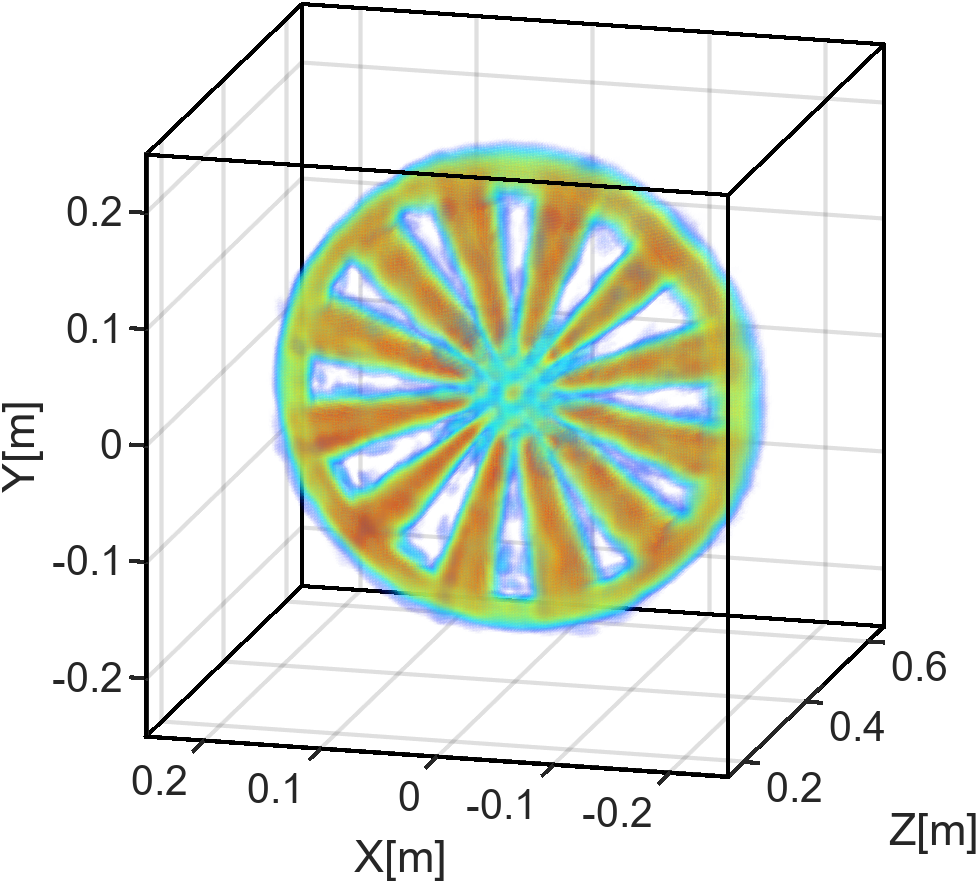}\\\footnotesize{(f)}
\end{minipage}
\begin{minipage}[t]{8.6cm}\centering
\includegraphics[width=6.0cm]{picture/color_bar.png}
\end{minipage} 
\caption{Imaging results of the different algorithms in Simulation \Rmnum{2}. (a)(c)(e) Maximum intensity projection results. (b)(d)(f) 3-D volume rendering results. (a)(b) Results of BPA. (c)(d) Results of EPC-RMA. (e)(f) Results of the proposed HHFFBPA.}
\label{fig_smssimimg}
\end{figure}

The imaging results of the expanded target simulation data with different algorithms are presented in Fig. \ref{fig_smssimimg}, and their PSNRs referenced to BPA are listed in Table \ref{tab_smssimimgpsnr}. Similar to the point target simulation, the proposed algorithm HHFFBPA achieves comparable image quality to BPA, while EPC-RMA suffers from defocusing and distortion in the squint region. Because the system parameters remained unchanged in both numerical simulations and the operations involved in the imaging algorithms are all insensitive to the target configurations, the computation time of different algorithms remains consistent with the results in Table \ref{tab_simalgotime}.

\begin{table}[htbp]\centering
\caption{Quantitative Metrics of the Imaging Results in Simulation \Rmnum{2}}
\begin{tabular}{
>{\raggedright\arraybackslash}p{2.5cm}
>{\raggedright\arraybackslash}p{2.5cm}}
\toprule
Algorithms  & PSNR [dB] \\
\midrule
BPA         & ----- \\
EPC-RMA     & 16.67 \\
HHFFBPA      & 36.33 \\
\bottomrule
\end{tabular}
\label{tab_smssimimgpsnr}
\end{table}

\subsection{Experimental Configuration}

\begin{figure}[!htbp]\centering
\includegraphics[width=8.6cm]{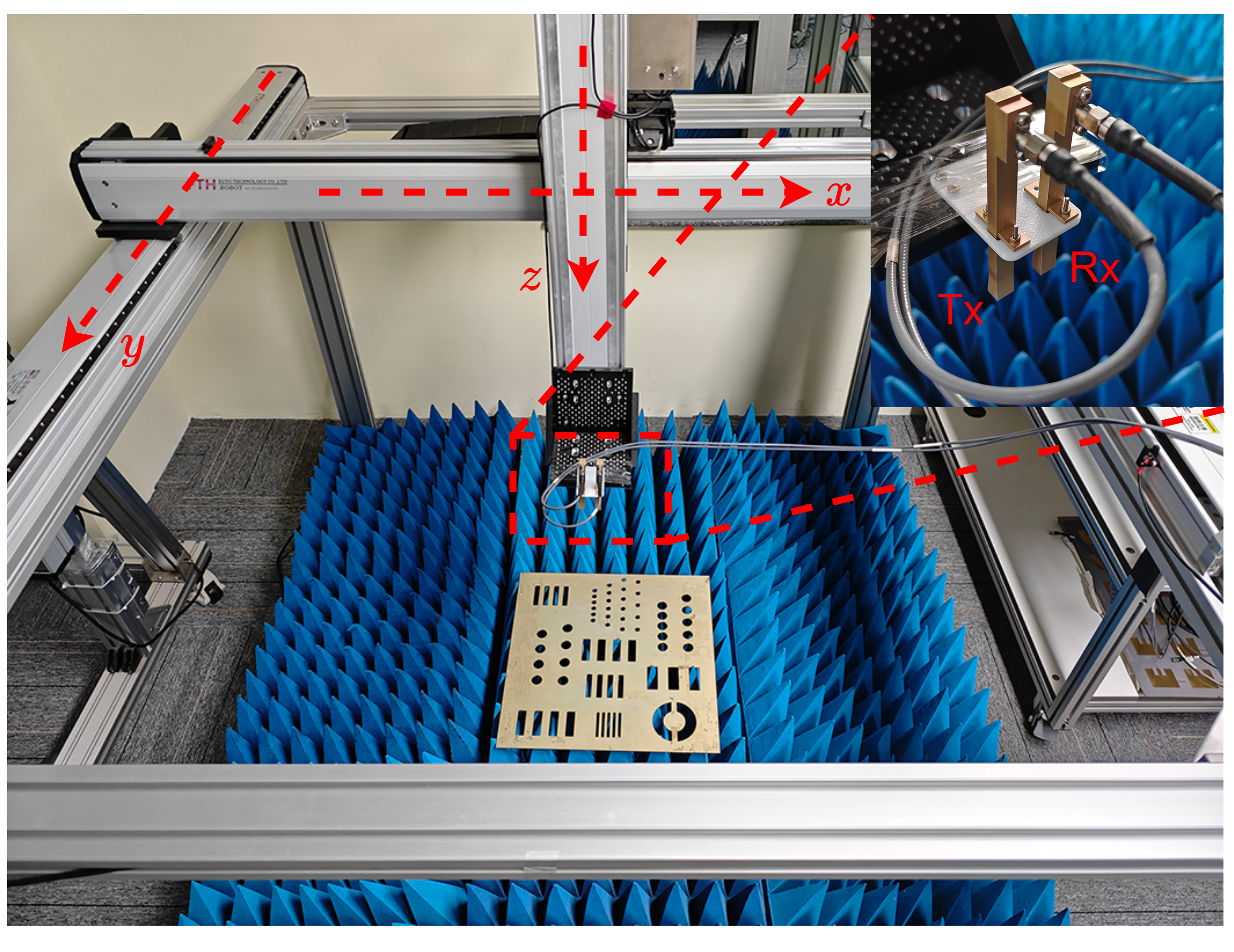}
\caption{Photograph of the test setup used in the experiments.}
\label{fig_testplatform} 
\end{figure}

The imaging performance of the proposed algorithm is further validated through experiments, with the experimental setup illustrated in Fig. \ref{fig_testplatform}. To simulate the trajectory fluctuations in handheld SAR applications, a 3-D mechanical scanning device is employed. The transmitting antenna and receiving antenna, each with a beamwidth of \SI{90}{\degree}, are fixed on the 3-D mechanical scanning device to enable radar measurements at the specified position. To maintain consistency with the numerical simulations, the array configuration remains unchanged for the experiments, as depicted in Fig. \ref{fig_arrpos}.
 
The remaining parameters are different from those in numerical simulations. The working frequency ranges from \SI{24.4}{\giga\hertz} to \SI{27.8}{\giga\hertz} with 64 equidistant samples. The dimensions of the imaging area are \SI{0.5}{\meter}$\times$\SI{0.5}{\meter}$\times$\SI{0.5}{\meter}, with 201$\times$201$\times$101 samples. The distance between the center of the imaging area and the center of the synthetic aperture is \SI{0.4}{\meter}.

\subsection{Experimental Results}

\begin{figure}[!htbp]\centering
\begin{minipage}[t]{3.5in}\centering
\includegraphics[width=3.5in]{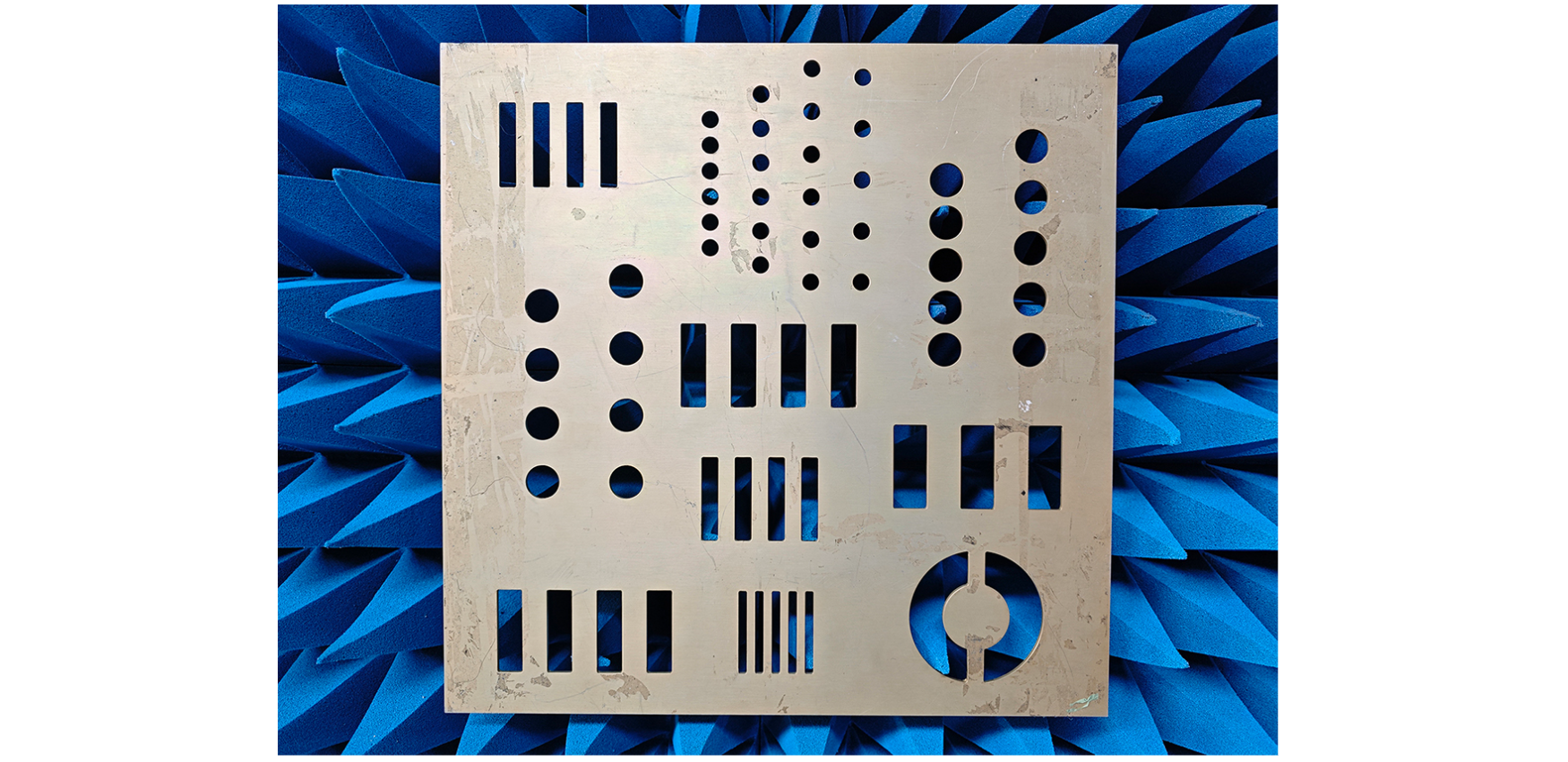}\\\footnotesize{(a)}
\end{minipage}
\begin{minipage}[t]{4.3cm}\centering
\includegraphics[width=4.3cm]{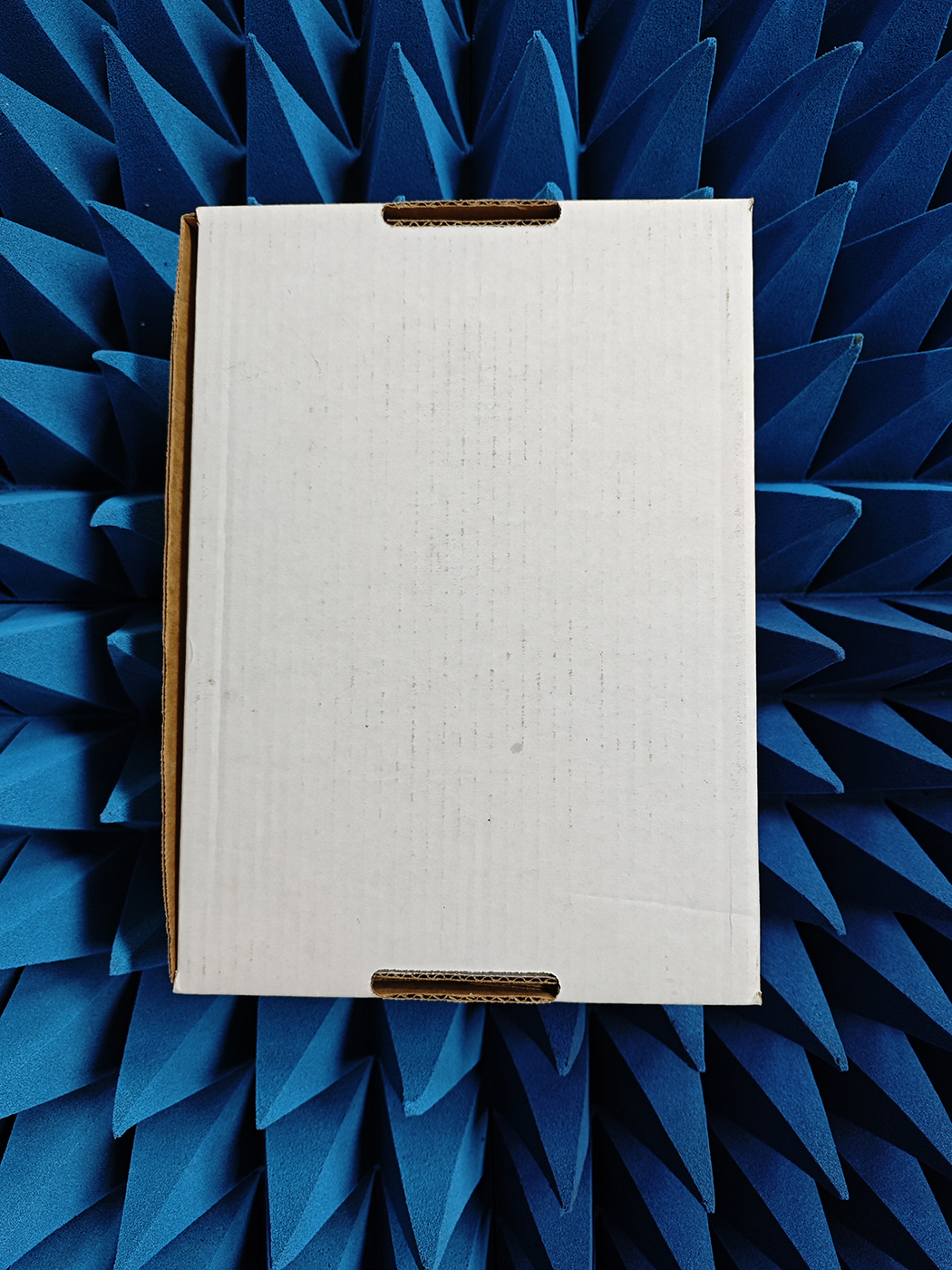}\\\footnotesize{(b)}
\end{minipage}
\begin{minipage}[t]{4.3cm}\centering
\includegraphics[width=4.3cm]{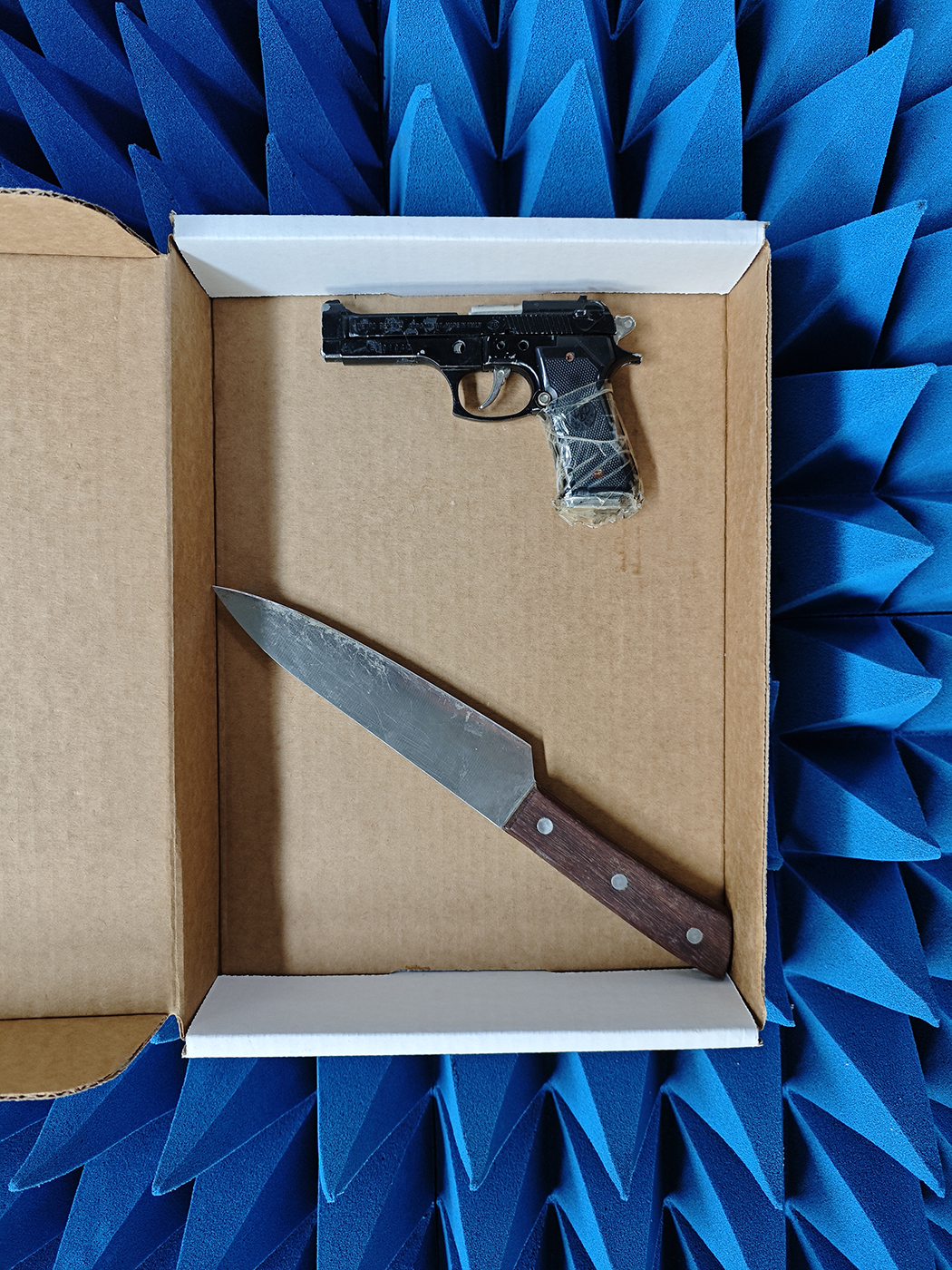}\\\footnotesize{(c)}
\end{minipage}
\caption{Imaging targets used in the experiments. (a) \SI{0.4}{m}$\times$\SI{0.4}{m} metal plate with various test patterns. (c) Cardboard box with a concealed model gun and knife inside. (b) Opened cardboard box to show the model gun and the knife inside.}
\label{fig_exptrg}
\end{figure}

To comprehensively demonstrate the imaging quality of the proposed algorithm, experiments are carried out with two distinct targets, as illustrated in Fig. \ref{fig_exptrg}. The first target is a \SI{0.4}{m}$\times$\SI{0.4}{m} metal plate with various test patterns, including round holes and slots of different sizes, to evaluate the spatial resolution of the imaging results. The other target consists of a concealed model gun and knife within a cardboard box, to demonstrate the imaging quality of the algorithms in real-world settings of concealed object detection. Both of the targets are positioned at a distance of \SI{0.4}{m} from the center of the synthetic aperture during their individual radar imaging measurements.

\begin{figure*}[!htbp]\centering
\begin{minipage}[t]{4.1cm}\centering
\includegraphics[width=4.1cm]{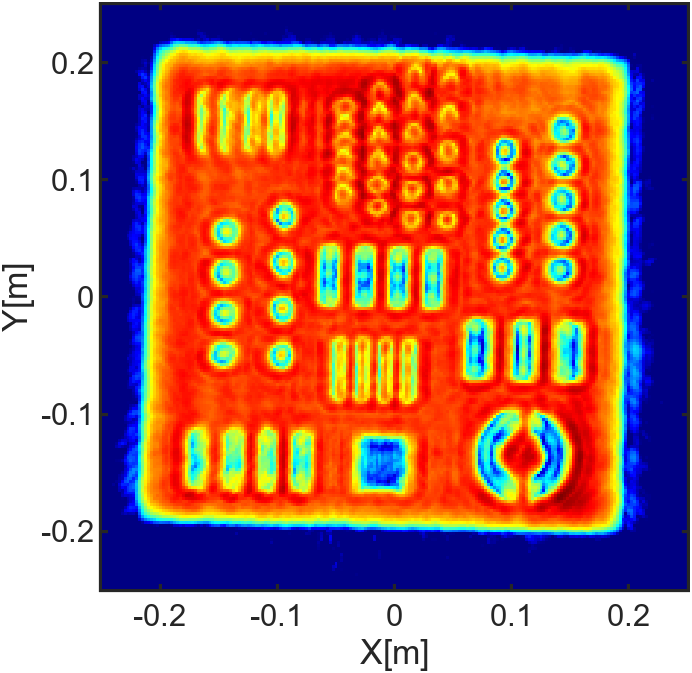}\\\footnotesize{(a)}
\end{minipage}
\begin{minipage}[t]{4.5cm}\centering
\includegraphics[width=4.5cm]{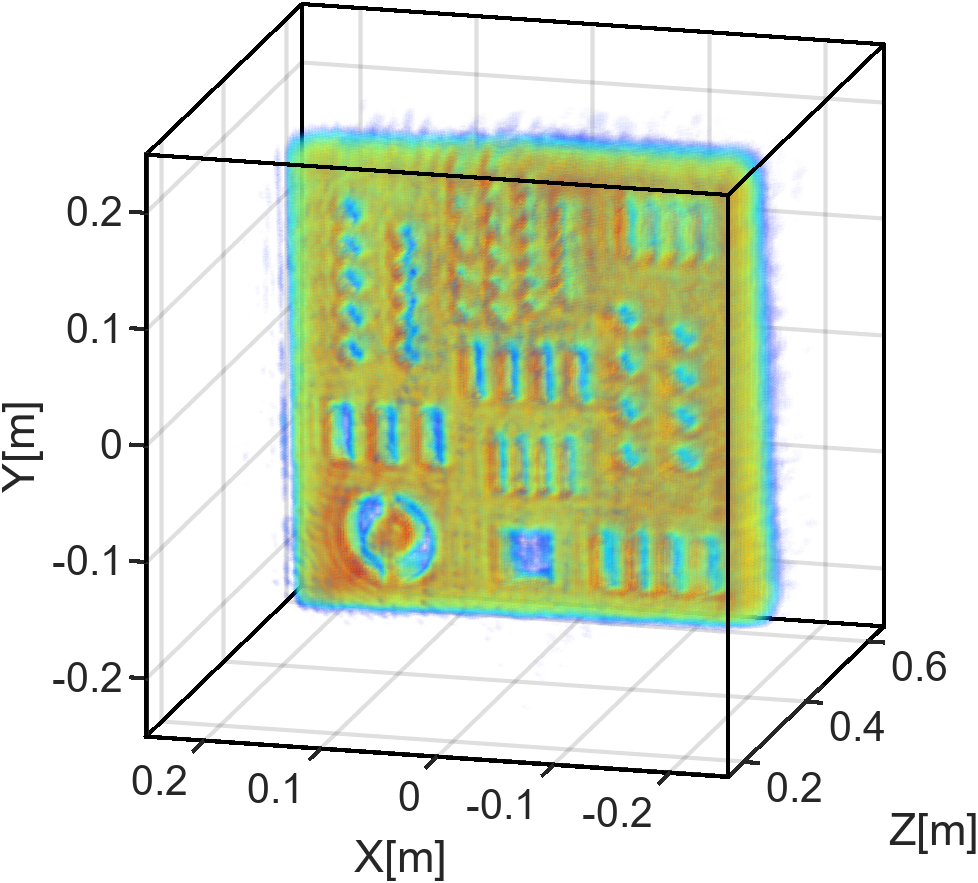}\\\footnotesize{(b)}
\end{minipage}
\begin{minipage}[t]{4.1cm}\centering
\includegraphics[width=4.1cm]{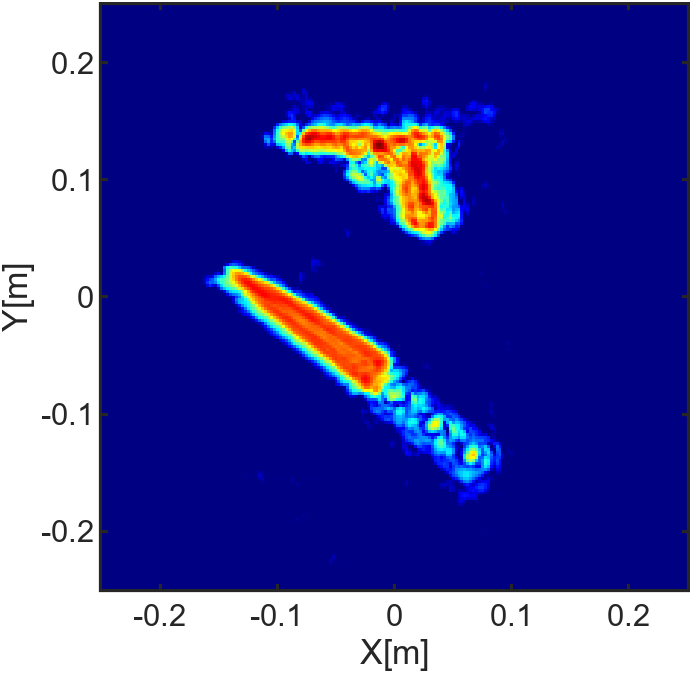}\\\footnotesize{(c)}
\end{minipage}
\begin{minipage}[t]{4.5cm}\centering
\includegraphics[width=4.5cm]{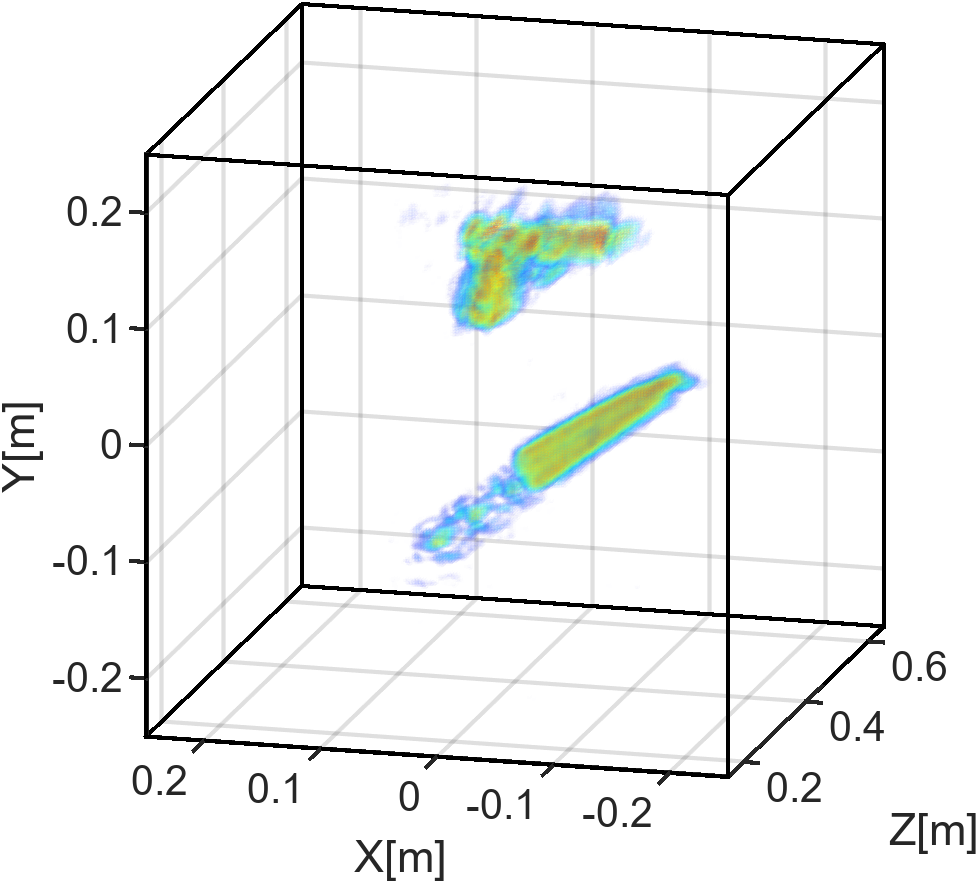}\\\footnotesize{(d)}
\end{minipage}
\begin{minipage}[t]{4.1cm}\centering
\includegraphics[width=4.1cm]{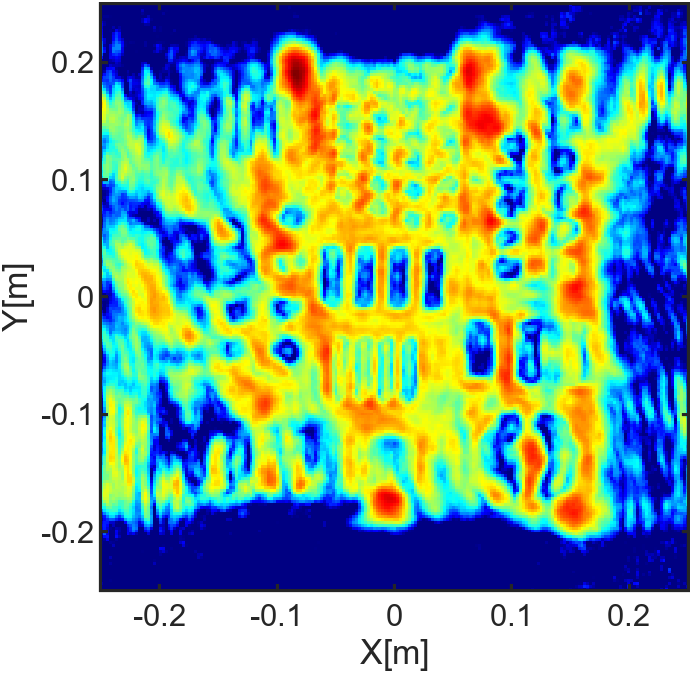}\\\footnotesize{(e)}
\end{minipage}
\begin{minipage}[t]{4.5cm}\centering
\includegraphics[width=4.5cm]{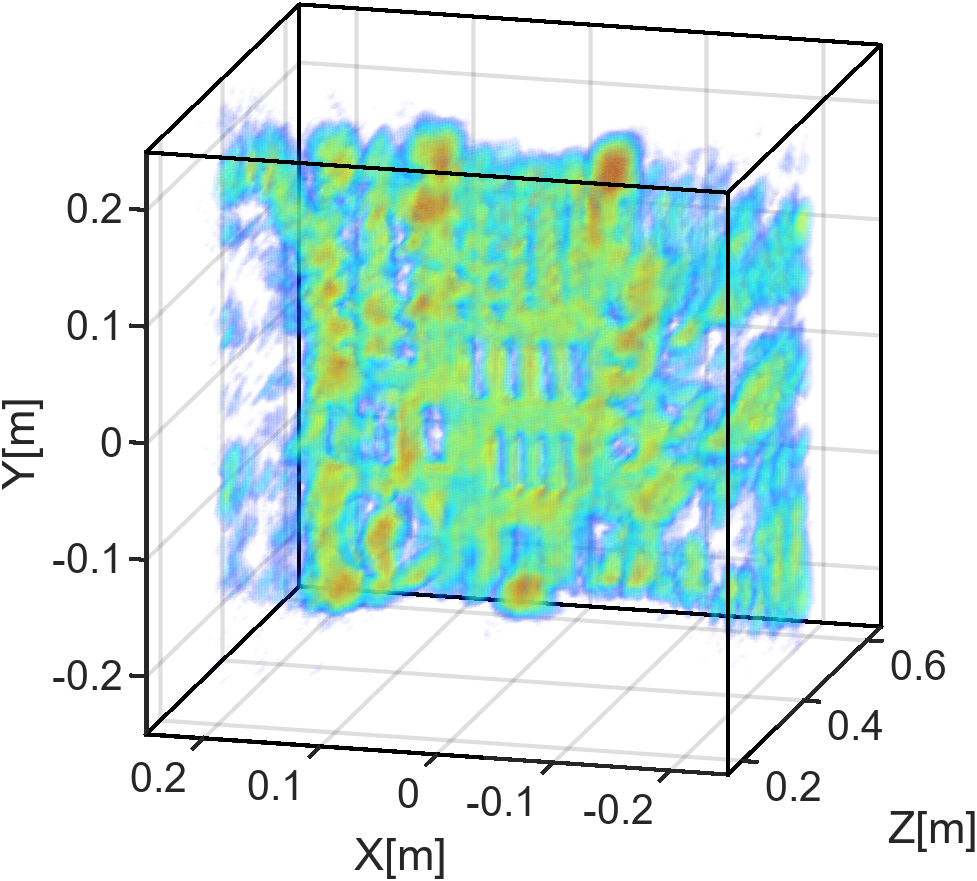}\\\footnotesize{(f)}
\end{minipage}
\begin{minipage}[t]{4.1cm}\centering
\includegraphics[width=4.1cm]{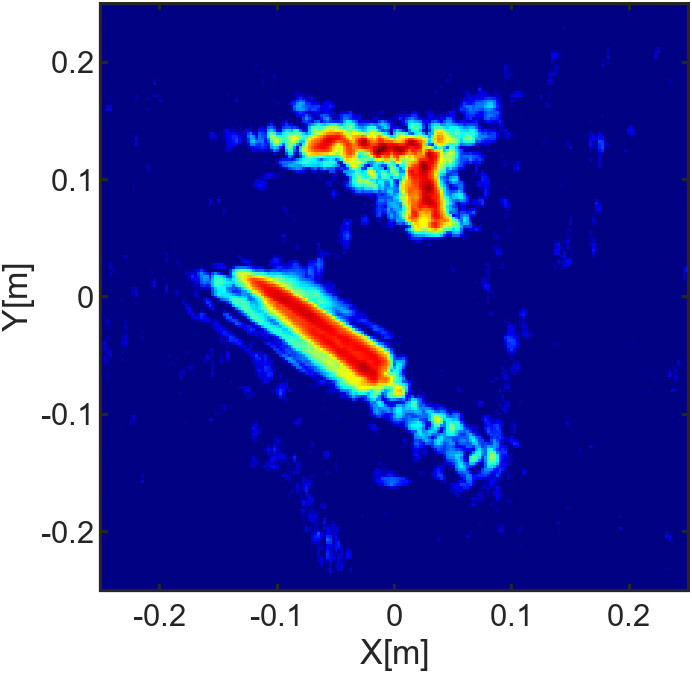}\\\footnotesize{(g)}
\end{minipage}
\begin{minipage}[t]{4.5cm}\centering
\includegraphics[width=4.5cm]{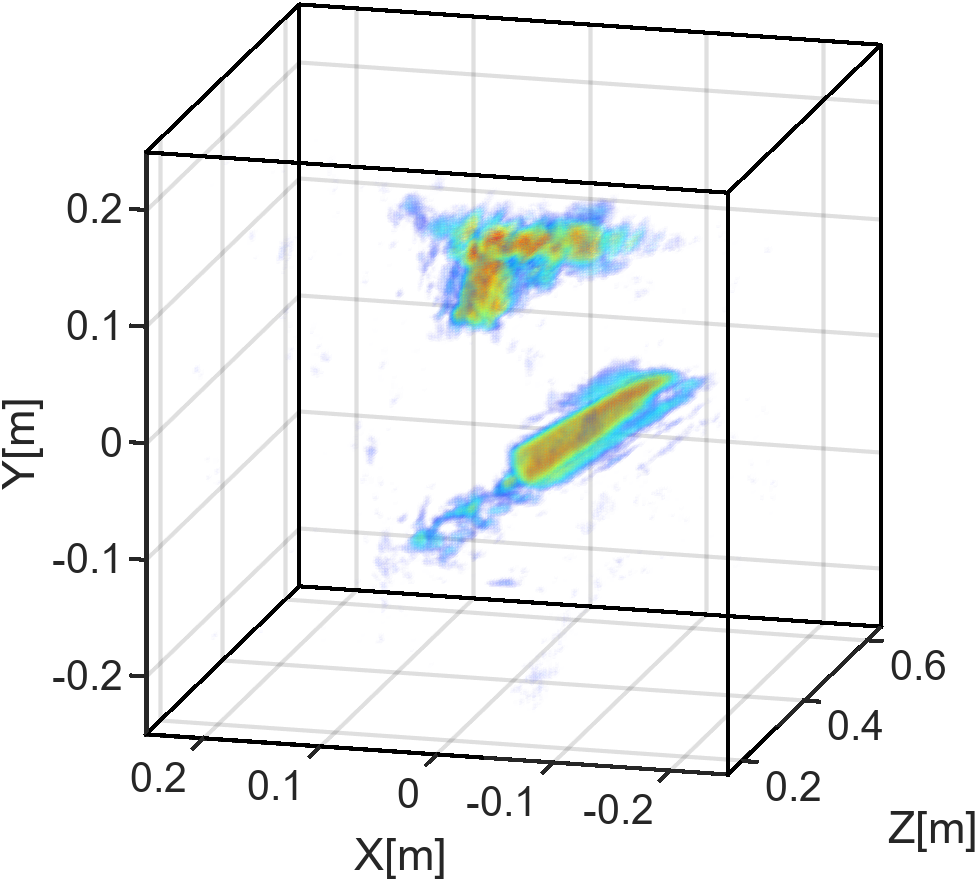}\\\footnotesize{(h)}
\end{minipage}
\begin{minipage}[t]{4.1cm}\centering
\includegraphics[width=4.1cm]{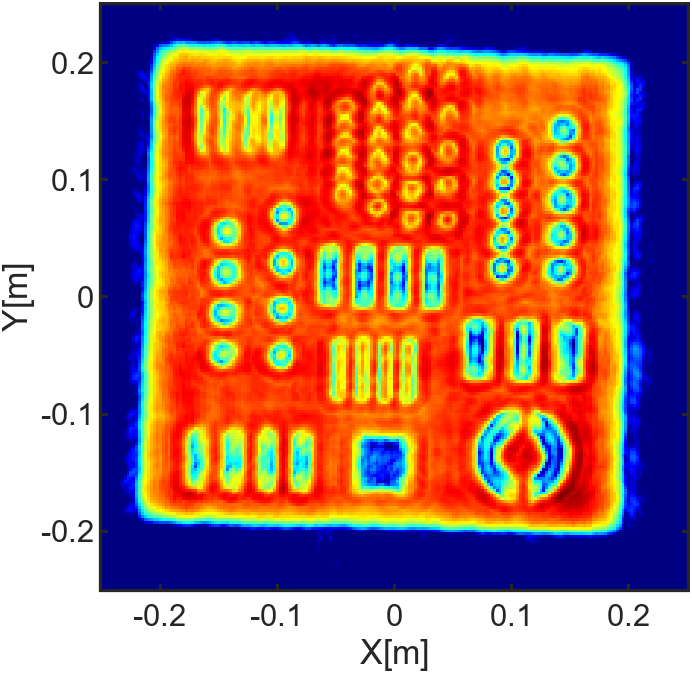}\\\footnotesize{(i)}
\end{minipage}
\begin{minipage}[t]{4.5cm}\centering
\includegraphics[width=4.5cm]{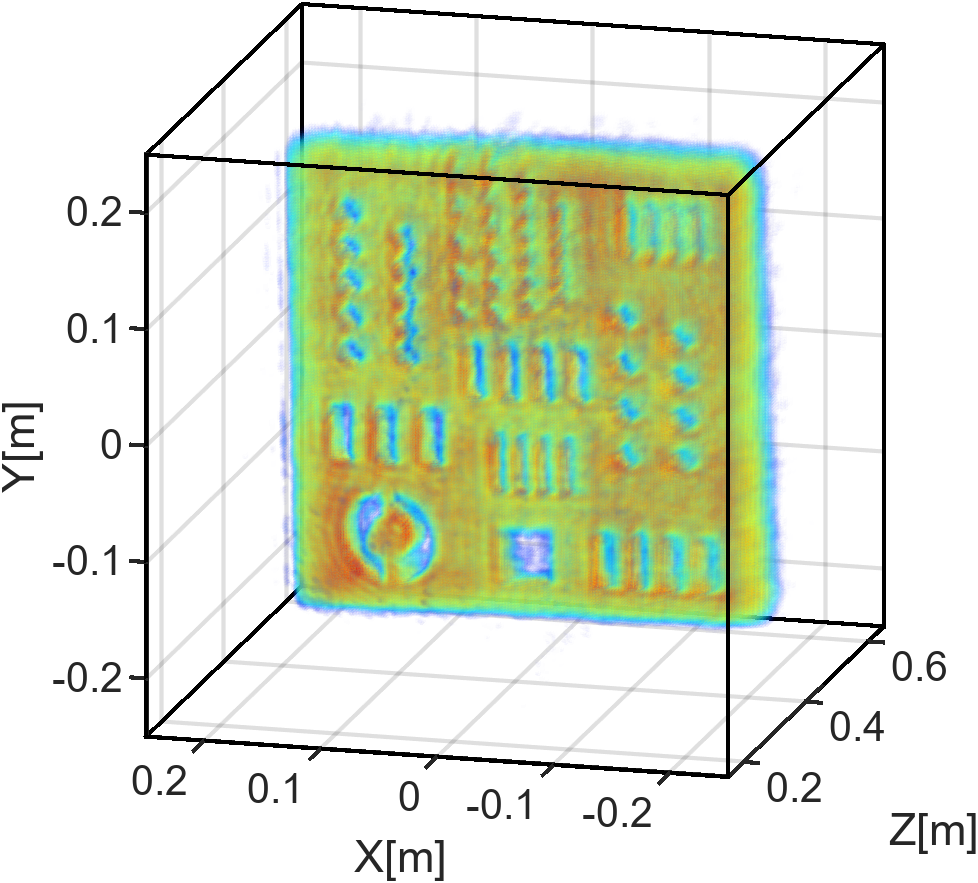}\\\footnotesize{(j)}
\end{minipage}
\begin{minipage}[t]{4.1cm}\centering
\includegraphics[width=4.1cm]{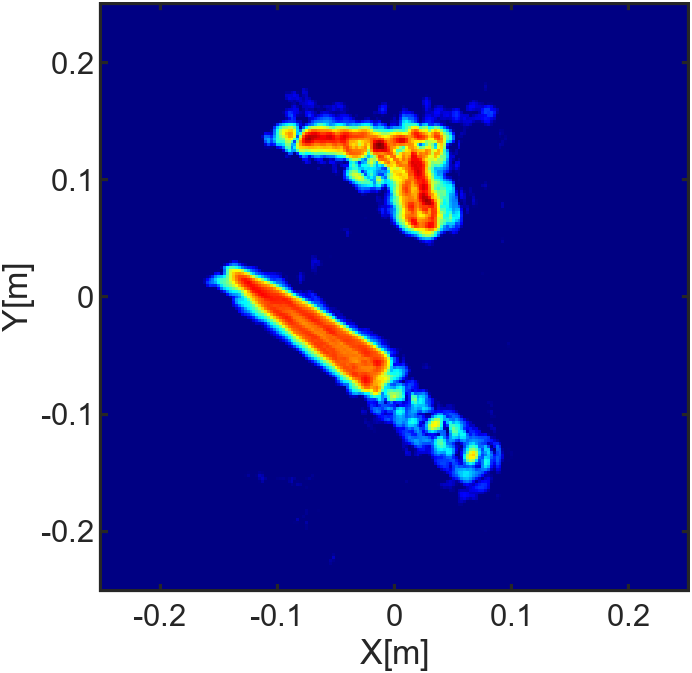}\\\footnotesize{(k)}
\end{minipage}
\begin{minipage}[t]{4.5cm}\centering
\includegraphics[width=4.5cm]{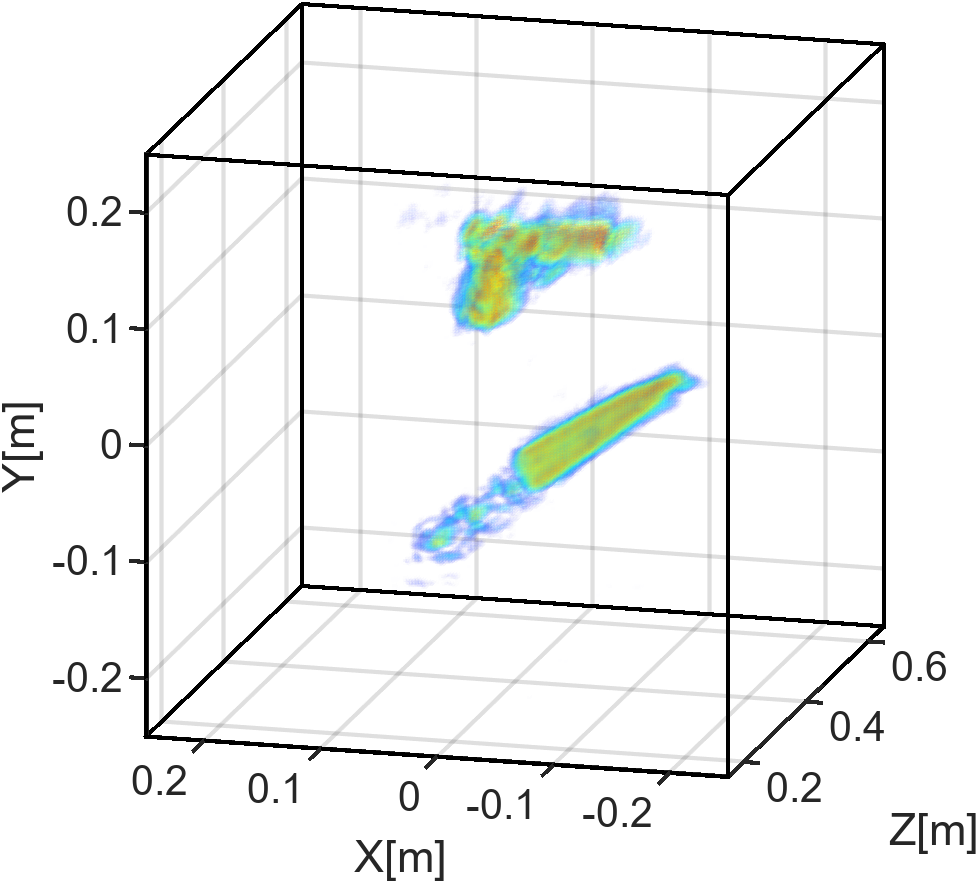}\\\footnotesize{(l)}
\end{minipage}
\begin{minipage}[t]{8.6cm}\centering
\includegraphics[width=6.0cm]{picture/color_bar.png}
\end{minipage} 
\caption{Imaging results of the different algorithms for different targets in the experiments. (a)(b)(e)(f)(i)(j) Results for the metal plate with various test patterns. (c)(d)(g)(h)(k)(l) Results for the cardboard box with concealed objects. (a)(c)(e)(g)(i)(k) Maximum intensity projection results. (b)(d)(f)(h)(j)(l) 3-D volume rendering results. (a)(b)(c)(d) Results of BPA. (e)(f)(g)(h) Results of EPC-RMA. (i)(j)(k)(l) Results of the proposed algorithm HHFFBPA.}
\label{fig_exptrgimg}
\end{figure*}

The imaging results of different algorithms for both targets are presented in Fig. \ref{fig_exptrgimg}, and their PSNRs relative to the results of BPA are listed in Table \ref{tab_exptrgimgidx}. Similar to the numerical simulations, the proposed algorithm HHFFBPA demonstrates comparable imaging quality to BPA with exceptional PSNRs of \SI{30.91}{dB} and \SI{44.59}{dB}. EPC-RMA exhibits good focusing capability at the center area but suffers from significant defocusing and distortion issues in the squint area, resulting in poor PSNRs of \SI{12.21}{dB} and \SI{26.04}{dB} respectively.

The computation time of different algorithms is also presented in Table \ref{tab_exptrgimgidx}. In comparison to the numerical simulations, a higher number of spatial samples of the imaging result is employed in experiments. As a result, the increase of the problem size $N$ causes a more significant contrast between the computation time of algorithms with varying computational complexities. The proposed algorithm HHFFBPA demonstrates an image reconstruction speed that is 99.26 times faster than BPA, which is higher than the 62.09 times improvement achieved in numerical simulations. Moreover, it is 10.14 times slower than EPC-RMA, which aligns closely with the 10.79 times observed in numerical simulations. The absolute time costs of the implemented algorithms are influenced by the variations in the numerical operation count and implementation-dependant execution efficiency of the algorithms. However, the scaling performance of the algorithms indicates that the proposed algorithm HHFFBPA achieves comparable computational complexities to EPC-RMA. 

\begin{table}[htbp]\centering
\caption{Quantitative Comparison of the Imaging Quality and The Computation Time of Different Algorithms in Experiments}
\begin{tabular}{
>{\raggedright\arraybackslash}m{1.5cm}
>{\raggedright\arraybackslash}m{1.5cm}
>{\raggedright\arraybackslash}m{1.5cm}
>{\raggedright\arraybackslash}m{1.5cm}}
\toprule
\multicolumn{1}{c}{\multirow{2}{*}{Algorithms}} & \multicolumn{2}{c}{PSNR {[}dB{]}} & \multicolumn{1}{c}{\multirow{2}{*}{Computation time {[}s{]}}} \\ 
\cline{2-3}
\multicolumn{1}{c}{} & Test patterns & Concealed objects & \multicolumn{1}{c}{} \\ 
\midrule
BPA     & ----- & ----- & 1489.85   \\
EPC-RMA & 12.21 & 26.04 & 1.48      \\
HHFFBPA & 30.91 & 44.59 & 15.01     \\ 
\bottomrule
\end{tabular}
\label{tab_exptrgimgidx}
\end{table}

Considering the above results on the imaging quality and the computational efficiency, the proposed algorithm HHFFBPA demonstrates comparable performance to BPA, while significantly outperforming EPC-RMA in terms of imaging quality, and exhibits equivalent computational complexity to EPC-RMA, thereby achieving exceptional overall imaging performance.

\section{Conclusion} \label{sec_conlu}

In this article, the fast factorized backprojection algorithm for near-range handheld synthetic aperture radar imaging is proposed. Unlike fast factorized time domain algorithms for far-field settings where the spectrum compression techniques of the subimages can be heuristically designed, the proposed algorithm analyzes the local spectral properties of the subimages for handheld SAR systems analytically and derives the effective spectrum compression techniques accordingly. Furthermore, the spectrum compression methods including SDC and LLT are formulated analytically and the algorithm can be performed directly without relying on numerically complex percalculations for the varying handheld synthetic aperture with arbitrary trajectory, achieving fast and accurate 3-D image reconstructions for handheld SAR systems. The imaging quality and computational efficiency of the proposed algorithm are verified with numerical simulations and experiments.

\bibliographystyle{IEEEtran}
\bibliography{reference.bib}

\newpage
 
\vspace{11pt}

\vfill

\end{document}